\documentclass[twocolumn,aps,prd,superscriptaddress,amsfont,amssymb,amsmath, showpacs,balancelastpage, nofootinbib]{revtex4-1}
\usepackage{graphicx,longtable,mathrsfs,color,array}
\usepackage{lmodern}

\usepackage[T1]{fontenc}
\usepackage[utf8]{inputenc}
\usepackage[USenglish]{babel}
\usepackage[hidelinks]{hyperref}
\usepackage{amssymb,amsmath,mathtools,mathrsfs}
\usepackage{epsfig,subfigure,placeins,float}
\usepackage[dvipsnames]{xcolor} 
\usepackage{booktabs,longtable,ctable,multirow}
\usepackage{exscale,relsize}
\usepackage[normalem]{ulem}
\usepackage{enumerate}
\usepackage{times, mathptmx}

\newcommand{\jcap}{JCAP}
\newcommand{\mnras}{MNRAS}

\newcommand{\hiclass}{{\tt hi\_class} }

\newcolumntype{C}[1]{>{\centering\let\newline\\\arraybackslash\hspace{0pt}}m{#1}}
\usepackage{colortbl} 
 
\usepackage{pifont} 
\usepackage{amsmath} 
\newcommand{\cmark}{\ding{51}}
\newcommand{\xmark}{\ding{55}}

\begin{document}
\title{A comparison of Einstein-Boltzmann solvers for testing General Relativity}

  \author{E.~Bellini}
  \affiliation{University of Oxford, Denys Wilkinson Building,
  	Keble Road, Oxford, OX1 3RH,  UK}
  \author{A.~Barreira}
  \affiliation{Max-Planck-Institut f\"{u}r Astrophysik, Karl-Schwarzschild-Str. 1, 85741 Garching, Germany}
 \author{ N.~Frusciante}
 \affiliation{Instituto de Astrofisica e Ciencias do Espaco, Faculdade de Ciencias da Universidade de Lisboa, Edificio C8, Campo Grande, P-1749016, Lisboa, Portugal}
 \author{B.~Hu}
 \affiliation{Department of Astronomy, Beijing Normal University, Beijing, 100875, China}
  \author{S.~Peirone}
  \affiliation{Institute Lorentz, Leiden University, PO Box 9506, Leiden 2300 RA, The Netherlands}
  \author{M.~Raveri} 
  \affiliation{Kavli Institute for Cosmological Physics, Enrico Fermi Institute,The University of Chicago, Chicago, Illinois 60637, USA}
  \author{M.~Zumalac\'arregui}
  \affiliation{Nordita, KTH Royal Institute of Technology and Stockholm University,
  	Roslagstullsbacken 23, SE-106 91 Stockholm, Sweden}
  \affiliation{Berkeley Center for Cosmological Physics, LBL and University of California at Berkeley, CA94720, USA}
 \author{A.~Avilez-Lopez}
 \affiliation{Departamento de F\'{\i}sica, Centro de Investigaci\'on y de Estudios
Avanzados del IPN, AP 14-740, Ciudad de M\'exico 07000, M\'exico} 
 \author{M.~Ballardini}
 \affiliation{Department of Physics \& Astronomy, University of the Western Cape, Cape Town 7535, South Africa}
 \affiliation{DIFA, Dipartimento di Fisica e Astronomia, Alma Mater Studiorum Università di Bologna, Viale Berti Pichat, 6/2, I-40127 Bologna, Italy}
 \affiliation{INAF/IASF Bologna, via Gobetti 101, I-40129 Bologna, Italy}
 \affiliation{INFN, Sezione di Bologna, Via Berti Pichat 6/2, I-40127 Bologna, Italy}
  \author{R.~A.~Battye}
  \affiliation{Jodrell Bank Centre for Astrophysics, School of Physics and Astronomy, The University of Manchester, 
Manchester, M13 9PL, U.K.}
  \author{B.~Bolliet}
  \affiliation{Laboratoire de Physique Subatomique et de Cosmologie, Universit\'e Grenoble-Alpes, CNRS/IN2P3\\
53, avenue des Martyrs, 38026 Grenoble cedex, France.}
 \author{E.~Calabrese}
 \affiliation{University of Oxford, Denys Wilkinson Building,
 	Keble Road, Oxford, OX1 3RH,  UK}
 \affiliation{School of Physics and Astronomy, Cardiff University, The Parade, Cardiff, CF24 3AA, UK}
\author{Y.~Dirian}
\affiliation{D\'{e}partement de Physique Th\'{e}orique and Center for Astroparticle Physics, Universit\'{e} de Gen\`{e}ve, 24 quai Ansermet, CH-1211 Gen\`{e}ve 4, Switzerland}
  \author{P.~G.~Ferreira}
  \affiliation{University of Oxford, Denys Wilkinson Building,
  	Keble Road, Oxford, OX1 3RH,  UK}
 \author{F.~Finelli}
 \affiliation{INAF/IASF Bologna, via Gobetti 101, I-40129 Bologna, Italy}
 \affiliation{INFN, Sezione di Bologna, Via Berti Pichat 6/2, I-40127 Bologna, Italy}
 \author{Z.~Huang}
 \affiliation{School of Physics and Astronomy, Sun Yat-sen University, 2 Daxue Road, Zhuhai, 519082, China}
\author{M.~M.~Ivanov}
\affiliation{Institute of Physics, LPPC, \'Ecole Polytechnique F\'ed\'erale de Lausanne, CH-1015, Lausanne, Switzerland}
\affiliation{Institute for Nuclear Research of the Russian Academy of Sciences,~60th October Anniversary Prospect, 7a, 117312
Moscow, Russia}
  \author{J.~Lesgourgues}
  \affiliation{Institute for Theoretical Particle Physics and Cosmology (TTK),
RWTH Aachen University, D-52056 Aachen, Germany.}
  \author{B.~Li}
 \affiliation{Institute for Computational Cosmology, Department of Physics, Durham University, Durham DH1 3LE, UK}
  \author{N.~A.~Lima}
  \affiliation{Institut f\"{u}r Theoretische Physik, Universit\"{a}t Heidelberg, Philosophenweg 16, D-69120 Heidelberg, Germany}
  \author{F.~Pace}
  \affiliation{Jodrell Bank Centre for Astrophysics, School of Physics and Astronomy, The University of Manchester, 
Manchester, M13 9PL, U.K.}
 \author{D.~Paoletti}
 \affiliation{INAF/IASF Bologna, via Gobetti 101, I-40129 Bologna, Italy}
 \affiliation{INFN, Sezione di Bologna, Via Berti Pichat 6/2, I-40127 Bologna, Italy}
  \author{I.~Sawicki}
  \affiliation{CEICO, Fyzik\'aln\'i ust\'av Akademie v\v{e}d \v{C}R, Na Slovance 1999/2, 182 21, Prague, Czechia}
  \author{A.~Silvestri}
  \affiliation{Institute Lorentz, Leiden University, PO Box 9506, Leiden 2300 RA, The Netherlands}
    \author{C.~Skordis}
  \affiliation{CEICO, Fyzik\'aln\'i ust\'av Akademie v\v{e}d \v{C}R, Na Slovance 1999/2, 182 21, Prague, Czechia}
  \affiliation{Department of Physics, University of Cyprus, 1, Panepistimiou Street, 2109, Aglantzia, Cyprus}
  \author{C.~Umilt\`{a}}
  \affiliation{Institut d'Astrophysique de Paris, CNRS (UMR7095), 98 bis Boulevard Arago, F-75014,
  	Paris, France}
  \affiliation{UPMC Univ Paris 06, UMR7095, 98 bis Boulevard Arago, F-75014, Paris, France}
  \affiliation{Sorbonne Universit\'es, Institut Lagrange de Paris (ILP), 98 bis Boulevard Arago, 75014
  	Paris, France}
  \author{F.~Vernizzi}
  \affiliation{Institut de Physique Th\' eorique, Universit\'e  Paris Saclay, 
CEA, CNRS, 91191 Gif-sur-Yvette, France}

\begin{abstract}
  We compare Einstein-Boltzmann solvers that include modifications to General Relativity and find that, for a wide range of models and parameters, they agree to a high level of precision. We look at three general purpose codes that primarily model general scalar-tensor theories, 
  three codes that model Jordan-Brans-Dicke (JBD) gravity,
  a code that models $f(R)$ gravity, 
  a code that models covariant Galileons, 
  a code that models Ho\v rava-Lifschitz gravity 
   and two codes that model non-local models of gravity. 
   Comparing predictions of the angular power spectrum of the cosmic microwave background and the power spectrum of dark matter for a suite of different models, we find agreement at the sub-percent level. This means that this suite of Einstein-Boltzmann solvers is now sufficiently accurate for precision constraints on cosmological and gravitational parameters. \end{abstract}

  \date{\today}
  \maketitle

\section{Introduction}
Parameter estimation has become an essential part of modern cosmology, e.g.~\cite{Ade:2015xua}. By this we mean the ability to constrain various properties of cosmological models using observational data such as the anisotropies of the cosmic microwave background (CMB), the large scale structure of the galaxy distribution (LSS), the expansion and acceleration rate of the universe and other such quantities. A crucial aspect of this endeavour is to be able to accurately calculate a range of observables from the cosmological models. This is done with Einstein-Boltzmann (EB) solvers, i.e. codes that solve the linearized Einstein and Boltzmann equations on an expanding background \cite{2003moco.book.....D}.

The history of EB solvers is tied to the success of modern theoretical cosmology. Beginning with the seminal work of Peebles and Yu \cite{1970ApJ...162..815P}, Wilson and Silk \cite{1981ApJ...243...14W}, Bond and Efstathiou \cite{1987MNRAS.226..655B} and Bertschinger and Ma \cite{1995ApJ...455....7M} these first attempts involved solving coupled set of many thousands of ordinary differential equations in a time consuming, computer intensive manner. A step change occurred with the introduction of the line of sight method and the CMBFAST code \cite{1996ApJ...469..437S} by Seljak and Zaldarriaga, which sped calculations up by orders of magnitude. Crucial in establishing the reliability of CMBFAST was a cross comparison \cite{Seljak:2003th} between a handful of EB solvers (including CMBFAST) that showed that it was possible to get agreement to within $0.1\%$. Fast EB solvers have become the norm:  CAMB \cite{2000ApJ...538..473L}, DASh \cite{Kaplinghat:2002mh}, CMBEASY \cite{2005JCAP...10..011D}  and CLASS \cite{Lesgourgues:2011re,Blas:2011rf} all use the line of sight approach and have been extensively used for cosmological parameters estimation. Of these, CAMB and CLASS are kept up to date and are, by far, the most widely used as part of the modern armoury of cosmological analysis tools.

While CAMB and CLASS were developed to accurately model the standard cosmology -- general relativity with a cosmological constant -- there has been surge in interest in testing extensions that involve modifications to gravity \cite{Clifton:2011jh}. Indeed, it has been argued that it should be possible to test general relativity (GR) and constrain the associated gravitational parameters to the same level of precision as with other cosmological parameters. More ambitiously, one hopes that it should be possible to test GR on cosmological scales with the same level of precision as is done on astrophysical scales \cite{Berti:2015itd}. Two types of codes have been developed for the purpose of achieving this goal: general purpose codes which are either not tied to any specific theory (such as MGCAMB \cite{Hojjati:2011ix} and ISITGR \cite{Dossett:2011tn} ) or model a broad class of (scalar-tensor) theories (such as \texttt{EFTCAMB} \cite{Hu:2013twa} and \hiclass \cite{Zumalacarregui:2016pph}) and specific codes which model targeted theories such as Jordan-Bran-Dicke gravity \cite{Avilez:2013dxa}, Einstein-Aether gravity \cite{Zuntz:2008zz}, $f(R)$ \cite{Bean:2006up}, covariant galileons \cite{Barreira:2012kk} and others.

The stakes have changed in terms of theoretical precision. Up and coming surveys such as 
 Euclid\footnote{{\tt https://www.euclid-ec.org/}}, LSST\footnote{\tt https://www.lsst.org/}, WFIRST\footnote{\tt https://wfirst.gsfc.nasa.gov/}, SKA\footnote{\tt http://skatelescope.org/} and Stage 4 CMB\footnote{\tt https://cmb-s4.org/} experiments all require sub-percent agreement in theoretical accuracy (cosmic variance is inversely proportional to the angular wavenumber probed, $\ell$, and we expect to at most, reach $\ell\sim$ few$\times 10^3$). While there have been attempts at checking and calibrating existing non-GR N-body codes \cite{Winther:2015wla}, until now the same effort has not been done for non-GR EB solvers with this accuracy in mind. In this paper we attempt to repeat what was done in \cite{Seljak:2003th,Lesgourgues:2011rg} with a handful of codes. We will focus on scalar modes, neglecting for simplicity primordial tensor modes and B-modes of the CMB. In particular, we will show that two general purpose codes -- \texttt{EFTCAMB} and \hiclass -- agree with each other to a high level of accuracy. The same level of accuracy is reached with the third general purpose code -- COOP; however, the latter code needs further calibration to maintain agreement at sub-Mpc scales. We also show that they agree with a number of other EB solvers for a suite of models such Jordan-Brans-Dicke (JBD), covariant Galileons, $f(R)$ and Ho\v{r}ava-Lifshitz (khronometric) gravity. And we will show that for some models not encompassed by these general purpose codes, i.e. non-local theories of gravity, there is good agreement between existing EB solvers targeting them. This gives us confidence that these codes can be used for precision constraints on general relativity using observables of a linearly perturbed universe.
 
We structure our paper as follows. In Section \ref{sec:theories} we layout the formalism used in constructing the different codes and we summarize the theories used in our comparison. In Section \ref{sec:codes} we describe the codes themselves, highlighting their key features and the techniques they involve. In Section \ref{sec:tests} we compare the codes in different settings. We begin by comparing the codes for specific models and then choose different families of parametrizations for the free functions in the general purpose codes. In Section \ref{sec:discussion} we discuss what we have learnt and what steps to take next in attempts at improving analysis tools for future cosmological surveys.

\section{Formalism and Theories}
\label{sec:theories}
To study cosmological perturbations on large scales, one must expand all relevant cosmological fields to linear order around a homogeneous and isotropic background. By cosmological fields we mean the space time metric, $g_{\mu\nu}$, the various components of the energy density, $\rho_i$ (where $i$ can stand for baryons, dark matter and any other fluid one might consider), the pressure, $P_i$, and momentum, $\theta_i$,  as well as the phase space densities of the relativistic components, $f_j$ (where $j$ now stands for photons and neutrinos) as well as any other exotic degree of freedom, (such as, for example, a  scalar field, $\phi$, in the case of quintessence theories). One then replaces these linearized fields in the cosmological evolution equations; specifically in the Einstein field equations, the conservation of energy momentum tensor and the Boltzmann equations. One can then evolve the background equations and the linearized evolution equations to figure out how a set of initial perturbations will evolve over time. 

The end goal is to be able to calculate a set of spectra. First, the power spectrum of matter fluctuations at conformal time $\tau$ defined by
\begin{eqnarray}
\langle \delta^*_{M}(\tau,{\bf k'})\delta_{M}(\tau,{\bf k})\rangle\equiv (2\pi)^3P(k,\tau)\delta^3({\bf k}-{\bf k'}) \;,
\end{eqnarray}
where we have expanded the energy density of matter, $\rho_M$ around its mean value, ${\bar \rho}_M$, $\delta_M=(\rho_M-{\bar \rho}_M)/{\bar \rho}_M$, and taken its Fourier transform. Second, the angular power spectrum of CMB anisotropies
\begin{eqnarray}
\langle a^*_{\ell'm'}a_{\ell m}\rangle=C^{TT}_\ell\delta_{\ell\ell'}\delta_{mm'} \;,
\end{eqnarray}
where we have expanded the anisotropies, $\delta T/T({\hat n})$ in spherical harmonics such that
\begin{eqnarray}
\frac{\delta T}{T}({\hat n})=\sum_{\ell m}a_{\ell m}Y_{\ell m}({\hat n})\;.
\end{eqnarray}
More generally one should also be able to calculate the angular power spectrum of polarization in the CMB, specifically of the "$E$" mode, $C^{EE}_\ell$, the "$B$" mode, $C^{BB}_\ell$ and the cross-spectra between the "$E$" mode and the temperature anisotropies, $C^{TE}_\ell$, as well as the angular power spectrum of the CMB lensing potential, $C^{\phi\phi}_\ell$.
As a by product, one  can also calculate "background" quantities such as the history of the Hubble rate, $H(\tau)$, the angular-distance as a function of redshift, $D_A(z)$ and other associated quantities such as the luminosity distance, $D_L(z)$.

To study deviations from general relativity, one needs to consider two main extensions. First one needs to include extra, gravitational degrees of freedom. In this paper we will restrict ourselves to scalar-tensor theories, as these have been the most thoroughly studied, and furthermore we will consider only one extra degree of freedom. This scalar field, and its perturbation, will have an additional evolution equation which is coupled to gravity. Second, there will be modifications to the Einstein field equations and their linearized form will be modified accordingly. How the field equations are modified and how the scalar field evolves depends on the class of theories one is considering. In what follows, we will describe  what these modifications mean for different classes of scalar-tensor theories and also theories that evolve restricted scalar degrees of freedom (such as Ho\v rava-Lifshitz and non-local theories of gravity).

\subsection{The Effective Field Theory of Dark Energy}\label{sec:eft}

A general approach to study scalar-tensor theories is the so-called Effective Field Theory of dark energy (EFT)~\cite{Gubitosi:2012hu,Bloomfield:2012ff,Gleyzes:2013ooa,Bloomfield:2013efa,Piazza:2013coa,Frusciante:2013zop,Gleyzes:2014rba,Gleyzes:2015pma,Perenon:2015sla,Kase:2014cwa,Linder:2015rcz,Frusciante:2016xoj}. Using this approach, it is possible to construct the most general action describing perturbations of single field dark energy (DE) and modified gravity models (MG). This can be done by considering all possible operators that satisfy spatial-diffeomorphism invariance, constructed from the metric in unitary gauge where the time is chosen to coincide with uniform field hypersurfaces. The operators can be ordered in number of perturbations and derivatives. Up to quadratic order in the perturbations, the action is given by
\begin{eqnarray}\label{eq:actionEFT}
  S &=& \int d^4x \sqrt{-g}   \left \{ \frac{M_{\rm Pl}^2}{2} [1+\Omega(\tau)]R+ \Lambda(\tau) - a^2c(\tau) \delta g^{00}\right.\nonumber \\
 &+&  \left.\frac{M_2^4 (\tau)}{2} (a^2\delta g^{00})^2 - \frac{\bar{M}_1^3 (\tau)}{2}a^2 \delta g^{00} \delta {K}^\mu_{\phantom{\mu}\mu} - \frac{\bar{M}_2^2 (\tau)}{2} (\delta K^\mu_{\phantom{\mu}\mu})^2 \right.
\nonumber \\ 
& - &\left.  \frac{\bar{M}_3^2 (\tau)}{2} \delta K^\mu_{\phantom{\mu}\nu} \delta K^\nu_{\phantom{\nu}\mu}
+\frac{a^2\hat{M}^2(\tau)}{2}\delta g^{00}\delta R^{(3)} \right.\nonumber \\
 & +& \left. m_2^2 (\tau) (g^{\mu \nu} + n^\mu n^\nu) \partial_\mu (a^2g^{00}) \partial_\nu(a^2 g^{00})
+ \ldots  \right\}\nonumber\\& +&  S_{m} [\chi_i ,g_{\mu \nu}],\nonumber\\
\end{eqnarray}
where $R$ is the 4D Ricci scalar and $n^\mu$ denotes the normal to the spatial hypersurfaces; $K_{\mu\nu} = (\delta^\rho_\mu + n^\rho n_\mu) \nabla_\rho n_\nu$ is the  extrinsic curvature, $K$ its trace, and $R^{(3)}$ is the 3D Ricci scalar, all defined with respect to the spatial hypersurfaces.
Moreover,  we have tagged with a $\delta$ all perturbations around the cosmological background. $S_m$ is the matter action describing the usual components of the Universe, which we assume to be minimally and universally coupled to gravity. The ellipsis stand for higher order terms that will not be considered here.
The  explicit evolution of the perturbation of the scalar field can be obtained by applying the  St\"{u}ckelberg  technique to Eq.~(\ref{eq:actionEFT}) which means restoring the time diffeomorphism invariance by an infinitesimal time coordinate transformation, i.e. $ t \rightarrow t + \, \pi(x^{\mu})$, where $\pi$ is the explicit scalar degree of freedom. 

In Eq.\ (\ref{eq:actionEFT}), the functions of time $\Lambda(\tau)$ and $c(\tau)$ can be expressed in terms of $\Omega(\tau)$, the Hubble rate and the matter background energy density and pressure, using the background evolution equations obtained from this action \cite{Gubitosi:2012hu,Bloomfield:2012ff,Gleyzes:2013ooa,Bloomfield:2013efa}. Then, the general family of scalar-tensor theories is spanned by eight functions of time, i.e.\ $\Omega(\tau)$, $M_2^4(\tau)$, $M^2_i(\tau)$ (with $i=1,\ldots,3$), ${\hat M}^2(\tau)$, $m^2_2(\tau)$ plus one function describing the background expansion rate as $H\equiv da/(adt)$.\footnote{Note that $H$ does not completely fix the evolution of all the background quantities; it must be augmented by the evolution of the matter species encoded in $S_m$.} Their time dependence is completely free unless they are constrained to represent some particular theory. Indeed, besides their model independent characterization, a general recipe exists to map specific models in the EFT language \cite{Gubitosi:2012hu,Bloomfield:2012ff,Bloomfield:2013efa,Gleyzes:2013ooa,Gleyzes:2014rba,Frusciante:2015maa,Frusciante:2016xoj}. In other words, by making specific choices for these EFT functions it is possible to single out a particular class of scalar-tensor theory and its cosmological evolution for a specific set of initial conditions. The number of EFT functions that are involved in the mapping increases proportionally to the complexity of the theory. In particular, linear perturbations in non-minimally coupled theories such as Jordan-Brans-Dicke are described in terms of two independent functions of time, $\Omega(\tau)$ and $H(\tau)$, i.e.\ by setting $M_2^4=0$, $\bar{M}^2_i=0$ ($i=1,\ldots,3$) and $m^2_2=0$. Increasing the complexity  of the theory, perturbations in Horndeski theories \cite{Horndeski:1974wa,Deffayet:2009mn} are described by setting $\{\bar{M}^2_2=-\bar{M}^2_3=2\hat{M}^2,m_2^2=0\}$, in which case one is left with four independent functions of time in addition to the usual dependence on $H(\tau)$ \cite{Gleyzes:2013ooa,Bloomfield:2013efa}. Moreover, by detuning $2\hat{M}^2$ from $\bar{M}^2_2=-\bar{M}^2_3$  one is considering beyond Horndeski theories \cite{Gleyzes:2014dya,Gleyzes:2014qga}. Lorentz violating theories, such as Ho\v rava gravity \cite{Horava:2008ih,Horava:2009uw}, also fall in this description by assuming $m_2^2\neq 0$.

For practical purposes, it is useful to define a set of dimensionless functions  in terms of the original EFT functions as
\begin{eqnarray}
\gamma_1&=&\frac{M^4_2}{M_{\rm Pl}^2H_0^2}\,, \ 
\gamma_2 =\frac{\bar{M}^3_1}{M_{\rm Pl}^2H_0}\,, \ 
\gamma_3 = \frac{\bar{M}^2_2}{M_{\rm Pl}^2}\,, 
\nonumber \\
\gamma_4 &=& \frac{\bar{M}^2_3}{M_{\rm Pl}^2}\,, \ \ \ \ \ \
\gamma_5 =\frac{\hat{M}^2}{M_{\rm Pl}^2}\,, \ \ \ \ \ \
\gamma_6  = \frac{m^2_2}{M_{\rm Pl}^2}\,,
\end{eqnarray}
where $H_0$ and $M_{\rm Pl}$ are the Hubble parameter today and the Planck mass respectively.

In this basis, Horndeski gravity corresponds to $\gamma_4=-\gamma_3$, $\gamma_5=\frac{\gamma_3}{2}$ and $\gamma_6=0$. As explained above, this reduces the number of free functions to five, i.e.\ $\{\Omega,\gamma_1,\gamma_2,\gamma_3\}$ plus a function that fixes the background expansion history. In this limit the EFT approach is  equivalent to the $\alpha$ formalism described in the next section. Indeed, a one-to-one map to convert between the two bases is provided in Appendix~\ref{sec:dictionary}.

\subsection{The Horndeski Action}

A standard approach to study general scalar-tensor theories is to write down a covariant action by considering explicitly combinations of a metric, $g_{\mu\nu}$, a scalar field, $\phi$, and their derivatives. The result for the most general action leading to second-order equations of motion on any background is the Horndeski action \cite{Horndeski:1974wa,Deffayet:2011gz}, which reads
\begin{eqnarray}\label{eq:L_horndeski}
S=\int d^4x \sqrt{-g}\sum_{i=2}^5{\cal L}_i[\phi,g_{\mu\nu}]+ S_{m} [\chi_i ,g_{\mu \nu}],
\end{eqnarray}
where, as always throughout this paper, we have assumed minimal and universal coupling to matter in $S_m$. The building blocks of the scalar field Lagrangian are
\begin{eqnarray}
{\cal L}_2&=& K ,  \nonumber \\
{\cal L}_3&=&  -G_3 \Box\phi , \nonumber \\
{\cal L}_4&=&   G_4R+G_{4X}\left\{(\Box \phi)^2-\nabla_\mu\nabla_\nu\phi \nabla^\mu\nabla^\nu\phi\right\}  , \nonumber \\
{\cal L}_5&=& G_5G_{\mu\nu}\nabla^\mu\nabla^\nu\phi
-\frac{1}{6}G_{5X}\big\{ (\Box\phi)^3
-3\nabla^\mu\nabla^\nu\phi\nabla_\mu\nabla_\nu\phi\Box\phi 
 \nonumber \\ & & 
+2\nabla^\nu\nabla_\mu\phi \nabla^\alpha\nabla_\nu\phi\nabla^\mu\nabla_\alpha \phi
\big\}   \,,
\label{eq:HorndeskiBB}
\end{eqnarray}
where $K$ and $G_A$ are functions of $\phi$ and $X\equiv-\nabla^\nu\phi\nabla_\nu\phi/2$, and the subscripts $X$ and $\phi$ denote derivatives. The four functions, $K$ and $G_A$ completely characterize this class of theories. 

Horndeski theories are not the most general viable class of theories. Indeed, it is possible to construct scalar-tensor theories with higher-order equations of motion and containing a single scalar degree of freedom, such as the so-called ``beyond Horndeski'' extension \cite{Zumalacarregui:2013pma,Gleyzes:2014dya,Gleyzes:2014qga}. It was recently realized that  higher-order scalar-tensor theories propagating a single scalar mode can be understood as degenerate theories \cite{Langlois:2015cwa,Crisostomi:2016czh,BenAchour:2016fzp}.

It is possible to prove that the exact linear dynamics predicted by the full Horndeski action, Eq.~(\ref{eq:L_horndeski}), is completely described by specifying five functions of time, the Hubble parameter and \cite{Bellini:2014fua}
\begin{eqnarray}
M^2_*&\equiv&2\left(G_4-2XG_{4X}+XG_{5\phi}-{\dot \phi}HXG_{5X}\right) , \nonumber \\
HM^2_*\alpha_M&\equiv&\frac{d}{dt}M^2_* , \nonumber \\
H^2M^2_*\alpha_K&\equiv&2X\left(K_X+2XK_{XX}-2G_{3\phi}-2XG_{3\phi X}\right) \nonumber \\ & &
+12\dot{\phi}XH\left(G_{3X}+XG_{3XX}-3G_{4\phi X}-2XG_{4\phi XX}\right) \nonumber \\ & &
+12XH^2\left(G_{4X}+8XG_{4XX}+4X^2G_{4XXX}\right)\nonumber \\ & &
-12XH^2\left(G_{5\phi}+5XG_{5\phi X}+2X^2G_{5\phi XX}\right)\nonumber \\ & &
+4\dot{\phi}XH^3\left(3G_{5X}+7XG_{5XX}+2X^2G_{5XXX}\right)\nonumber , \\
HM^2_*\alpha_B&\equiv&2\dot{\phi}\left(XG_{3X}-G_{4\phi}-2XG_{4\phi X}\right) \nonumber \\ & &
+8XH\left(G_{4X}+2XG_{4XX}-G_{5\phi}-XG_{5\phi X}\right)  \nonumber \\ & &
+2\dot{\phi}XH^2\left(3G_{5X}+2XG_{5XX}\right) \nonumber , \\ 
M^2_*\alpha_T&\equiv&2X\left[2G_{4X}-2G_{5\phi}-\left(\ddot{\phi}-\dot{\phi}H\right)G_{5X}\right] \label{eq:alphas}\,,
\end{eqnarray}
where dots are derivatives w.r.t.\ cosmic time $t$ and $H\equiv da/(adt)$.

While the Hubble parameter fixes the expansion history of the universe, the $\alpha_i$ functions appear only at the perturbation level. $M_*^2$ defines an effective Planck mass, which canonically normalize the tensor modes. $\alpha_K$ and $\alpha_B$ (dubbed as \textit{kineticity} and \textit{braiding}) are respectively the standard kinetic term present in simple DE models such as quintessence and the kinetic term arising from a mixing between the scalar field and the metric, which is typical of MG theories as $f(R)$. Finally, $\alpha_T$ has been named \textit{tensor speed excess}, and it is responsible for deviations on the speed of gravitational waves while on the scalar sector it generates anisotropic stress between the gravitational potentials.

It is straightforward to relate the free functions $\{M_*, \alpha_K, \alpha_B,\alpha_T\}$ defined above to the free functions $\{\Omega, \gamma_1, \gamma_2, \gamma_3\}$ used to describe Horndeski theories in the EFT formalism. The  mapping between these sets of functions is reported in Appendix~\ref{sec:dictionary}. For an explicit expression of the functions $\{\Omega, \gamma_1, \gamma_2, \gamma_3\}$ in terms of the original $\{K, G_A\}$ in Eq.\ (7), we refer the reader to \cite{Frusciante:2016xoj} (see also \cite{Gleyzes:2013ooa,Bloomfield:2013efa}).

Regardless of the basis ($\alpha$s or EFT), it is clear now that there are two possibilities. The first one is to calculate the time dependence of $\alpha_i$ or $\gamma_i$ and the background consistently to reproduce a specific sub-model of Horndeski, the second one is to specify directly their time dependence.
Finally, the evolution equation for the extra scalar field and the modifications to the gravitational field equations depend solely on this set of free functions; any cosmology arising from Horndeski gravity can be modelled with an appropriate time dependence for these free functions.

\subsection{Jordan-Brans-Dicke}
The Jordan-Brans-Dicke (JBD) theory of gravity \cite{Brans:1961sx}, a particular case of the Horndeski theory, is given by the action
\begin{eqnarray}
S=\int d^4x \sqrt{-g}\frac{M_{\rm Pl}^2}{2}\left[\phi R-\frac{\omega_{\rm BD}}\phi\nabla_\mu\phi\nabla^\mu\phi-2V\right]+ S_{m} [\chi_i ,g_{\mu \nu}]\,, \nonumber \\
\end{eqnarray}
where $V(\phi)$ is a potential term and $\omega_{\rm BD}$ is a free parameter. GR is recovered when $\omega_{\rm BD}\rightarrow \infty$. For our test, we will not consider a generic potential but a cosmological constant instead, $\Lambda$, as the source of dark energy.

In the EFT language, linear perturbations in JBD theories are described by two functions, i.e. the Hubble rate $H(t)$ (or equivalently $c(\tau)$ or $\Lambda(\tau)$) and
\begin{align}\label{EFT_JBD}
&\Omega(\tau)=\phi-1 \,, \nonumber \\
&\gamma_i(\tau)=0\,.
\end{align}
We can see that in this case there are no terms consisting of purely modified perturbations (i.e.\ any of the $\gamma_i$).

Alternatively the $\alpha_i(\tau)$ functions read
\begin{eqnarray}
\alpha_M(\tau)&=&\frac{d\ln \phi}{d\ln a}, \nonumber \\
\alpha_B(\tau)&=&-\alpha_M, \nonumber \\
\alpha_K(\tau)&=&\omega_{\rm BD}\alpha^2_M, \nonumber \\
\alpha_T(\tau)&=&0. \label{eq:JBDalphas}
\end{eqnarray}
As with the EFT basis, one has to consider the Hubble parameter $H(\tau)$ as an additional building function. However, $H(\tau)$ can be written entirely as a function of the $\alpha$s, meaning that the five functions of time needed to describe the full Horndeski theory reduce to two in the JBD case, consistently with the EFT description of the previous paragraph.

In order to fix the above functions one has to solve the background equations to determine the time evolution of $\{H, \phi\}$.  

\subsection{Covariant Galileon}
\label{sec:galileons_theory}
The covariant Galileon model corresponds to the subclass of scalar-tensor theories of Eq.~(\ref{eq:L_horndeski}) that (in the limit of flat spacetime) is invariant under a {\it Galilean shift} of the scalar field \cite{Nicolis:2008in}, i.e.\ $\partial_\mu\phi\rightarrow \partial_\mu\phi+b_\mu$ (where $b_\mu$ is a constant four-vector). The covariant construction of the model presented in \cite{Deffayet:2009wt} consists in the addition of counter terms that cancel higher-derivative terms that would otherwise be present in the naive covariantization (i.e.\ simply replacing partial with covariant derivatives; see however \cite{Gleyzes:2014dya} for why the addition of these counter terms is not strictly necessary). Galilean invariance no longer holds in spacetimes like FRW, but the resulting model is one with a very rich and testable cosmological behaviour. The Horndeski functions in Eqs.~(\ref{eq:HorndeskiBB}) have this form
\begin{eqnarray}
{\cal L}_{2} & = & c_{2}X -\frac{c_{1}M^{3}}{2}\phi \,,\label{eq:L2}\\
{\cal L}_{3} & = & 2\frac{c_{3}}{M^{3}}X\Box\phi\,,\label{eq:L3}\\
{\cal L}_{4} & = & \left(\frac{M_{p}^{2}}{2}+\frac{c_{4}}{M^{6}}X^{2}\right)R
+2\frac{c_{4}}{M^{6}}X\left[\left(\Box\phi\right)^{2}-\phi_{;\mu\nu}\phi^{;\mu\nu}\right]\,,\label{eq:L4}\\
{\cal L}_{5} & = & \frac{c_{5}}{M^{9}}X^{2}G_{\mu\nu}\phi^{;\mu\nu}
-\frac{1}{3}\frac{c_{5}}{M^{9}}X\left[\left(\Box\phi\right)^{3}+2{\phi_{;\mu}}^{\nu}{\phi_{;\nu}}^{\alpha}{\phi_{;\alpha}}^{\mu} \right. \nonumber \\   & & \left. \ \ \ \ -3\phi_{;\mu\nu}\phi^{;\mu\nu}\Box\phi\right]\,,\label{eq:L5}
\end{eqnarray}
Here, as usual, we have set $M^{3}=H_{0}^{2}M_{p}$. Note that these definitions are related to Ref. \cite{Barreira:2014jha} by $c_3^{\rm ours}\to -c_3^{\rm theirs}$ and $c_5^{\rm ours} = 3c_5^{\rm theirs} $. 
There is some freedom to rescale the field and normalize some of the coefficients. Following Ref. \cite{Barreira:2014jha} we can choose $c_2<0$ and rescale the field so that $c_2 = -1$ (models with $c_2>0$ have a stable Minkowski limit with $\phi_{,\mu}=0$ and thus no acceleration without a cosmological constant, see e.g.\ \cite{Brax:2014vla}). The term proportional to $\phi$ in $\mathcal{L}_2$ is uninteresting, so we will set $c_1 = 0$ from now on. This leaves us with three free parameters, $c_{3,4,5}$.

An analysis of Galileon cosmology was undertaken in \cite{DeFelice:2010pv,Barreira:2014jha} identifying some of the key features which we briefly touch upon. The Galileon contribution to the energy density at $a=1$ is \cite{DeFelice:2010pv}
\begin{equation}\label{eq:Omega_de}
\Omega_{gal}=-\frac{1}{6}\xi^{2}-2c_{3}\xi^{3}+\frac{15}{2}c_{4}\xi^{4}+\frac{7}{3}c_{5}\xi^{5}\,,
\end{equation}
(defined such that the coefficients are dimensionless) and where
\begin{equation}
\xi\equiv \frac{\dot{\phi}H}{M_{\rm Pl} H_{0}^{2}}\,.
\end{equation}

Given that the theory is shift symmetric, there is an associated Noether current satisfying $\nabla_\mu J^\mu = 0$ \cite{Deffayet:2010qz}. For a cosmological background $ J^i = 0$, $J^0\equiv n$ and the shift-current decays with the expansion $n \propto a^{-3} \to0$ at late times. The field evolution is thus driven to an attractor where
\begin{equation}\label{eq:attractor}
J^0 \propto -\xi-6c_{3}\xi^{2}+18c_{4}\xi^{3}+5c_{5}\xi^{4} =0\,,
\end{equation}
i.e.\ $\xi$ is a constant and the evolution of the background is independent of the initial conditions of the scalar field.
Although it has been claimed that background observations favour a non-scaling  behaviour of the scalar field \cite{Neveu:2016gxp}, CMB observations (not considered in Ref.~\cite{Neveu:2016gxp}) require that the tracker has been reached before Dark Energy dominates (Fig.~11 of Ref.~\cite{Barreira:2014jha}).\footnote{Note that if inflation occurred it would set the field very near the attractor by the early radiation era \cite{Deffayet:2010qz,Renk:2017rzu}.} So if only considering the evolution on the attractor, one can use Eqs.~(\ref{eq:Omega_de},\ref{eq:attractor}) to trade two of the independent $c_i$ for $\xi$ and $\Omega_{gal}$. 

It has thus become standard to refer to three models:
\begin{enumerate}
 \item Cubic: $c_4=c_5=0$, with $c_3$ the only free parameter; choosing $\Omega_{gal}$ determines determes $\xi$. No additional parameters compared to $\Lambda$CDM.
 
 \item Quartic: $c_5 = 0$; $\Omega_{gal}$ and $\xi$ are free parameters. One more parameter than $\Lambda$CDM.
 \item Quintic:  $c_3, \xi, \Omega_{gal}$ are free parameters. Two extra parameters relative to $\Lambda$CDM.
\end{enumerate}
All of these models are self-accelerating models without a cosmological constant, and hence do not admit a continuous limit to $\Lambda$CDM.

The covariant Galileon model is implemented in \texttt{EFTCAMB} and GALCAMB assuming the attractor solution Eq.~(\ref{eq:attractor}); on the other hand \hiclass solves the the full background equations both on- and off-attractor. The two approaches are equivalent if ones chooses the initial conditions for the scalar field on the attractor, which will be the strategy for the rest of the Galileon comparison. When the attractor solution is considered with the above conventions,  the alpha functions read

\begin{eqnarray}
 M_*^2\alpha_{\rm K} {\cal E}^4 &=& - \xi^2-12 c_3 \xi^3+54 c_4 \xi^4+20 c_5 \xi^5 \,,\nonumber
 \\
 M_*^2\alpha_{\rm B} {\cal E}^4 &=& -2 c_3 \xi^3+12 c_4 \xi^4+5 c_5 \xi^5 \,,\nonumber
\\
 M_*^2\alpha_{\rm M} {\cal E}^4 &=&  6 c_4 \frac{\dot{{\cal H}}}{{\cal H}^2} \xi^4+4 c_5 \frac{\dot{{\cal H}}}{{\cal H}^2} \xi^5 \,,\nonumber
 \\
 M_*^2\alpha_{\rm T} {\cal E}^4 &=& 2 c_4 \xi^4+c_5 \xi^5\left(1+\frac{\dot{{\cal H}}}{{\cal H}^2}\right) \,,
 \end{eqnarray}
where ${\cal E} = {\cal H}(\tau)/{ H}_0$ is the dimensionless expansion rate with ${\cal H}=aH$ and a dot now denotes a derivative w.r.t.\ conformal time, $\tau$. 
With the same conventions, the EFT functions read
\begin{eqnarray}
\Omega &=& \frac{a^4 H_0^4 \xi ^4 \left(\mathcal{H}^2 \left(c_4-2 c_5 \xi \right)+2 c_5 \xi  \dot{\mathcal{H}}\right)}{2 \mathcal{H}^6} \,,\nonumber\\
\gamma_3 &=&-\frac{a^4 H_0^4 \xi ^4 \left(2 c_4 \mathcal{H}^2+ c_5 \xi  \dot{\mathcal{H}}\right)}{\mathcal{H}^6}\,,\nonumber\\
\gamma_2 &=&-\frac{a^3 H_0^3 \xi ^3}{\mathcal{H}^7}\Bigg[ c_5 \xi ^2 \mathcal{H} \ddot{\mathcal{H}}+2 \xi  \mathcal{H}^2 \left(4 c_5 \xi -c_4\right) \dot{\mathcal{H}}\nonumber\\ &&+\mathcal{H}^4 \left(\xi  \left( c_5 \xi
   +14 c_4\right)-2 c_3\right)-6 c_5 \xi ^2 \dot{\mathcal{H}}^2\Bigg]\,, \nonumber\\
\gamma_1 &=&\frac{a^2 H_0^2 \xi ^3}{4 \mathcal{H}^8}\Bigg[2 \xi  \mathcal{H}^3 \left(5 c_5 \xi -c_4\right) \ddot{\mathcal{H}} +42 c_5 \xi ^2 \dot{\mathcal{H}}^3 \\&&+\mathcal{H}^4 \left(9 \xi  \left(\frac{7}{3} c_5 \xi -2 c_4\right)+2
  c_3\right) \dot{\mathcal{H}}\nonumber\\ &&+\xi  \mathcal{H}^2 \left( c_5 \xi  \dddot{\mathcal{H}}+10 \left(c_4-5 c_5 \xi \right) \dot{\mathcal{H}}^2\right)\nonumber\\ &&-18 c_5 \xi ^2 \mathcal{H} \dot{\mathcal{H}}
   \ddot{\mathcal{H}}+4 \mathcal{H}^6 \left(3 \xi  \left( c_5 \xi +4 c_4\right)-2 c_3\right)\Bigg]\nonumber\,.
\end{eqnarray}


\subsection{f(R) gravity}
\label{sec:f(R)}
$f(R)$ models of gravity are described by the following Lagrangian in the Jordan frame
\begin{equation}\label{action_fR}
S=\int d^4x \sqrt{-g} \left[R+f(R)\right]+S_m[\chi_i,g_{\mu\nu}] \,,
\end{equation}
where $f(R)$ is a generic function of the Ricci scalar and the matter fields $\chi_i$ are minimally coupled to gravity. 
They represent a popular class of scalar-tensor theories which has been extensively studied in the literature~\cite{Song:2006ej,Bean:2006up,Amendola:2006kh,Pogosian:2007sw,DeFelice:2010aj} and for which N-body simulation codes exist~\cite{Zhao:2010qy,Baldi:2013iza,Lombriser:2011zw,Hammami:2015iwa,Winther:2015wla}. Depending on the choice of the functional form of $f(R)$, it is possible to design models that obey stability conditions and give a viable cosmology~\cite{Amendola:2006kh,Sawicki:2007tf,Pogosian:2007sw}. A well-known example of viable model that also obeys solar system constraints is the one introduced by Hu $\&$ Sawicki in~\cite{Hu:2007nk}.

The higher order  nature of the theory, offers an alternative way of treating $f(R)$ models, i.e.\ via the so-called \emph{designer} approach. In the latter, one fixes the expansion history and uses the Friedmann equation as a second-order differential equation for $f[R(a)]$ to reconstruct the $f(R)$ model corresponding to the chosen history~\cite{Song:2006ej,Pogosian:2007sw}. Generically, for each expansion history,  one finds a family of viable models that reproduce it and are commonly labelled by the boundary condition at present time, $f_R^0$. Equivalently, they can be parametrized by the present day value of the function
\begin{equation}
\label{ComptonWave}
B=\frac{f_{RR}}{1+f_R}R^\prime\frac{H}{H^\prime}\,,
\end{equation}
where a prime denotes derivation w.r.t. $\ln{a}$. The smaller the value of $B_0$, the smaller the scale at which the fifth force introduced by $f(R)$ kicks in.
As in the JBD case, $f(R)$ models are described in the EFT formalism by two functions \cite{Gubitosi:2012hu}, the Hubble parameter and
\begin{eqnarray}\label{fR_matching}
&& \Omega=f_R\nonumber\\
&&\gamma_i(\tau)=0\,.
\end{eqnarray}
This has been used to implement $f(R)$ gravity into \texttt{EFTCAMB}, both for the designer models as well as for the Hu-Sawicki one~\cite{Hu:2013twa,Hu:2016zrh}. Alternatively, they can be described by the Equation of State approach (EoS) implemented in {\tt CLASS\_EOS\_fR}~\cite{Battye:2013aaa,Battye:2015hza}.

In this comparison we will focus on designer $f(R)$ models, since our aim is that of comparing the Einstein-Boltzmann solvers at the level of their predictions for linear perturbations. 

\subsection{Hořava-Lifshitz gravity}
\label{sec:HLtheory}

This model was introduced in Ref.~\cite{Horava:2009uw}. It was extended in Ref.~\cite{Blas:2010hb}, where it was shown that action for the low-energy \textit{healthy} version of Ho\v rava-Lifshitz gravity is given by 
\begin{eqnarray}\label{eq:actionhorava}
\mathcal{S}_{H}&=&\frac{1}{16\pi G_H}\int{}d^4x\sqrt{-g}[ K_{ij}K^{ij}-\lambda K^2 -2 \xi\bar{\Lambda}  
\nonumber \\ & +&\xi R^{(3)}+\eta a_i a^i] +S_m[\chi_i,g_{\mu\nu}]\,,
\end{eqnarray}
where $\lambda$, $\eta$, and $\xi$ are dimensionless coupling constants,  $\bar{\Lambda}$ is the ``bare'' cosmological constant and $G_H$ is the ``bare" gravitational constant related to Newton's constant via $1/16\pi G_H=M_{\rm Pl}^2/(2\xi-\eta)$~\cite{Blas:2009qj}.
Note that the choice $\lambda=\xi=1,\eta=0$ restores GR. 
In general, departures from these values lead to the violation of the local Lorentz symmetry of GR
and the appearance of a new scalar degree of freedom, known as the \textit{khronon}.
It should be pointed out that the model \eqref{eq:actionhorava} is equivalent to khronometric gravity \cite{Blas:2009qj},
an effective field theory which explicitly operates the khronon.\footnote{In turn, khronometric gravity is a variant 
of Einstein--Aether gravity \cite{Jacobson:2000xp},
an effective field theory describing the effects of Lorentz invariance violation. It should be pointed out that these models have identical scalar and tensor sectors.}
The correspondence between $\{\lambda, \eta, \xi \}$ and the coupling constants of the khronometric model 
$\{\alpha, \beta, \lambda\}$ 
is 
\begin{eqnarray}
\eta= -\frac{\alpha_{kh} }{\beta_{kh} -1},\quad \xi = -\frac{1}{\beta_{kh} -1},\quad \lambda = -\frac{{\lambda}_{kh} +1}{\beta_{kh} -1}\,,
\end{eqnarray}
where the subscript $kh$ is added for clarity.

The parameters $\lambda$, $\eta$, and $\xi$ are subject to various constraints from 
the absence of the vacuum Cherenkov radiation,
Solar system tests, astrophysics, and cosmology \cite{Elliott:2005va,Blas:2009qj,Jacobson:2000xp,Blas:2010hb,Yagi:2013ava,Blas:2016qmn,Audren:2014hza,Frusciante:2015maa}. 
The cosmological consequences of this model have been investigated in Refs.~\cite{ArmendarizPicon:2010rs,Kobayashi:2010eh,Blas:2012vn,Blas:2011en,Audren:2013dwa},
including interesting phenomenological implications for dark matter and dark energy.

The map of the action Eq.~(\ref{eq:actionhorava}) to the EFT functions~\cite{Frusciante:2015maa} is 
\begin{eqnarray}\label{eq:Horava_mapping}
\Omega&=&\frac{\eta}{(2\xi-\eta)}, \nonumber \\
\gamma_4&=& -\frac{2}{(2\xi-\eta)}(1-\xi),\nonumber \\
\gamma_3 &=&-\frac{2}{(2\xi-\eta)}(\xi-\lambda), \nonumber \\
\gamma_6&=&\frac{\eta}{4(2\xi-\eta)},\nonumber \\
\gamma_1&=&\frac{1}{2a^2H_0^2(2\xi-\eta)}(1+2\xi-3\lambda)\left(\dot{{\cal H}}-{\cal H}^2\right) \,,\nonumber\\
\gamma_2&=&\gamma_5=0,
\end{eqnarray}
which has been implemented in \texttt{EFTCAMB}~\cite{Hu:2014oga}.

\subsection{Non-local gravity}
\label{sec:NL-theory}
\def\nloc{RR}
\def\lcdm{\Lambda{\rm CDM}}
The non-local theory we consider here is that put forward in \cite{Maggiore:2014sia} (known as the $\nloc$ model for short), which is described by the action
\begin{eqnarray}\label{eq:NLG-action}
S_{\nloc} = \frac{1}{16\pi G}\int {\rm d}^4 x\sqrt{-g}\left[R - \frac{m^2}{6}R\Box^{-2}R - \mathcal{L}_M\right],
\end{eqnarray}
where $\mathcal{L}_M$ is the Lagrange density of minimally coupled matter fields and $\Box^{-1}$ is a formal inverse of the d'Alembert operator $\Box = \nabla^\mu\nabla_\mu$. The latter can be expressed as, 
\begin{eqnarray}\label{eq:formalgreen}
(\Box^{-1}A)(x) = A_{\rm hom}(x) - \int{\rm d}^4y\sqrt{-g(y)}G(x, y)A(y),
\end{eqnarray}
where $A$ is some scalar function of the spacetime coordinate $x$, and the homogeneous solution $A_{\rm hom}(x)$ and the Green's function $G(x, y)$ specify the definition of the $\Box^{-1}$ operator. Eq.~(\eqref{eq:NLG-action}) is meant to be understood as a toy-model to explore the phenomenology of the $R\Box^{-2}R$ term, while a deeper physical motivation for its origin is still not available (see \cite{Maggiore:2016gpx} and references therein for works along these lines). In the absence of such a fundamental understanding, different choices for the structure of the $\Box^{-1}$ operator (i.e.\ different homogeneous solutions and $G(x, y)$) should be regarded as different non-local models altogether, and the mass scale $m$ treated as a free parameter.

In cosmological studies of the $\nloc$ model, it has become common to cast the action of Eq.~(\ref{eq:NLG-action}) into the following "localized" form
\begin{eqnarray}\label{eq:NLG-action-local}
S_{\nloc, \mathrm{loc}}~ &&= \frac{1}{16\pi G}\int {\rm d}^4 x\sqrt{-g}\left[R - \frac{m^2}{6}R S - \xi_1\left(\Box U + R\right) \right. \nonumber \\
&&\ \ \ \ \ \ \ \ \ \ \ \ \  \ \ \ \ \ \ \ \ \ \ \ \ \ \ \ \ \ \ \ \left. - \xi_2\left(\Box S + U\right) - \mathcal{L}_m\right],
\end{eqnarray}
where $U$ and $S$ are two auxiliary scalar fields and $\xi_1$ and $\xi_2$ are two Lagrange multipliers that enforce the constraints
\begin{eqnarray}
\label{eq:u}\Box U &=& - R , \ \ \ \ \ \ \\
\label{eq:s}\Box S &=& - U  .
\end{eqnarray}
Invoking a given (left) inverse, one can solve the last two equations formally as
\begin{eqnarray}
\label{eq:iu} U &=& - \Box^{-1} R , \ \ \ \ \ \ \\
\label{eq:is} S &=& - \Box^{-1} U = \Box^{-2} R.
\end{eqnarray}
This allows one to integrate out $U$ and $S$ from the action (as well as $\xi_{1}$ and $\xi_{2}$), thereby recovering the original non-local action. The equations of motion associated with the action of Eq.~(\ref{eq:NLG-action-local}) are 
\begin{eqnarray}
&&\label{eq:fe1}G_{\mu\nu} - \frac{m^2}{6}K_{\mu\nu} = {8 \pi G} T_{\mu\nu}, \\
&&\label{eq:fe2}\Box U = -R, \\
&&\label{eq:fe3}\Box S = -U,
\end{eqnarray}
with
\begin{eqnarray}
\label{eq:fe4}K_{\mu\nu} \equiv 2SG_{\mu\nu} - 2\nabla_\mu\nabla_\nu S - 2\nabla_{(\mu}S\nabla_{\nu)}U \nonumber \\ 
+ \left(2\Box S + \nabla_\alpha S\nabla^\alpha U - \frac{U^2}{2}\right)g_{\mu\nu}.
\end{eqnarray}
An advantage of using Eq.~(\ref{eq:NLG-action-local}) is that the resulting equations of motion become a set of coupled differential equations, which are comparatively easier to solve than the integro-differential equations of the non-local version of the model. To ensure causality one must impose by hand that the Green's function used within $\Box^{-1}$ in Eqs.~(\ref{eq:iu}) and (\ref{eq:is}) is of the retarded kind and this condition is naturally satisfied in integrating the localized version forward in time. Further, the quantities $U$ and $S$ should not be regarded as physical propagating scalar degrees of freedom, but instead as mere auxiliary scalar functions that facilitate the calculations. In practice, this means that once the homogeneous solution associated with $\Box^{-1}$ is specified, then the differential equations of the localized problem must be solved with the one compatible choice of initial conditions of the scalar functions. Here, we fix $U$, $S$ and their first derivatives to zero, deep in the radiation dominated regime (this is as was done, for instance, in \cite{Dirian:2016puz, 2014JCAP...09..031B}; see \cite{2016arXiv160604349N} for a study of the impact of different initial conditions) which corresponds to choosing vanishing homogeneous solutions for them. Once the initial conditions of the $U$ and $S$ scalars are fixed, then the only remaining free parameter in the model is the mass scale $m$, which effectively replaces the role of $\Lambda$ in $\lcdm$ and can be derived from the condition to render a spatially flat Universe.

Finally, note that the Horndeski Lagrangian is a local theory featuring one propagating scalar degree of freedom, and hence, does not encompass the $\nloc$ model.
\section{The Codes} \label{sec:codes}
\begin{figure*}[t!]
\includegraphics[width=0.9\textwidth]{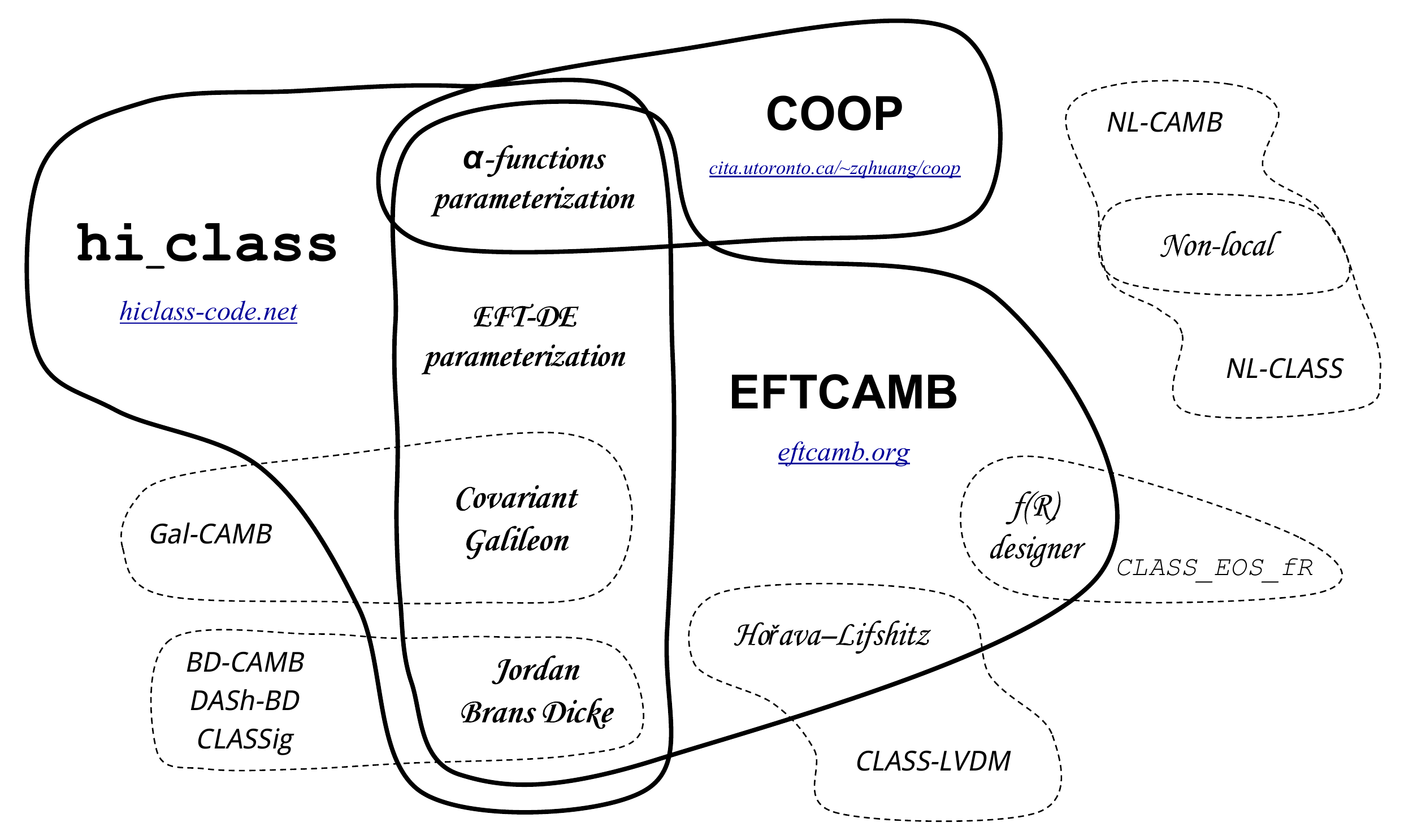}
\vspace*{-3mm}
\caption{Overlap between codes and theories used in the comparison. Each code is represented by a silhouette that covers the models for which it has been compared. General-purpose and publicly available codes are represented by thick solid regions, while model-specific or private codes are enclosed by dashed lines. Note that we only show the models used in this paper, not the full theory space available to each code.}\label{fig:code_theory_overview}
\end{figure*}
There are a number of EB solvers, some of which are described below, developed to explore deviations of GR. 
While, schematically, we have summarized how to study linear cosmological perturbations, there are a number of subtleties which we will mention now briefly. For a start, there is redundancy (or gauge freedom) in how to parametrize the scalar modes of the linearized metrics; typically EB solvers make a particular choice of gauge -- the synchronous gauge -- although another common gauge -- the Newtonian gauge -- is particularly useful in extracting physical understanding of the various effects at play. Also it should be noted that the universe undergoes an elaborate thermal history: it will recombine and subsequently reionize. It is essential to model this evolution accurately as it has a significant effect on the evolution of perturbations. Another key aspect is the use of line of sight methods (mentioned in the introduction) that substantially speed up the numerical computation of the evolution of perturbations by many orders of magnitude; as shown in \cite{1996ApJ...469..437S} it is possible to obtain an accurate solution of the Boltzmann hierarchy by first solving a truncated form of the lower order moments of the perturbation variables and then judiciously integrating over the appropriate kernel convolved with these lower order moments. All current EB solvers use this approach. 

Most (but not all) EB solvers currently being used are modifications of either CAMB or CLASS. This means that they have evolved from very different code bases, are in different languages and use (mostly) different algorithms. This is of tremendous benefit when we compare results in the next section. We should highlight, however, that there are a couple of cases -- DASh and COOP -- that do not belong to this genealogy.

The codes used in this comparison, along with the models tested, are summarized in Fig.~\ref{fig:code_theory_overview} and Tab.~\ref{tab:code_theory_overview} and the details of each code can be found in the following sections.

\begin{table*}
\begin{tabular}{cccccccc}
&  $\alpha$& EFT & JBD & Covariant & f(R)&Ho\v rava &Non-Local \\ 
& Parametrization & Parametrization &  & Galileon & designer & Lifshitz& Gravity\\ 
\hline
\rowcolor[gray]{.8}\texttt{EFTCAMB}&\cmark&\cmark&\cmark&\cmark&\cmark&\cmark&\xmark\\
\rowcolor[gray]{.9}\texttt{hi\_class}&\cmark&\cmark&\cmark&\cmark& \xmark&\xmark &\xmark\\
\rowcolor[gray]{.8}COOP&\cmark&\xmark&\xmark&\xmark&\xmark&\xmark&\xmark\\
\rowcolor[gray]{.9}GalCAMB&\xmark&\xmark&\xmark&\cmark&\xmark&\xmark&\xmark\\
\rowcolor[gray]{.8}BD-CAMB&\xmark&\xmark&\cmark&\xmark&\xmark&\xmark&\xmark\\
\rowcolor[gray]{.9}DashBD&\xmark&\xmark&\cmark&\xmark&\xmark&\xmark&\xmark\\
\rowcolor[gray]{.8}CLASSig&\xmark&\xmark&\cmark&\xmark&\xmark&\xmark&\xmark\\
\rowcolor[gray]{.9}{\tt CLASS\_EOS\_fR}&\xmark&\xmark&\xmark&\xmark&\cmark&\xmark&\xmark\\
\rowcolor[gray]{.8}CLASS-LVDM&\xmark&\xmark&\xmark&\xmark&\xmark&\cmark&\xmark\\
\rowcolor[gray]{.9}NL-CLASS&\xmark&\xmark&\xmark&\xmark&\xmark&\xmark&\cmark\\
\rowcolor[gray]{.8}NL-CAMB&\xmark&\xmark&\xmark&\xmark&\xmark&\xmark&\cmark\\
\end{tabular}
\caption{We show schematically the codes used in this comparison along with the models tested. This Table provides the same information as Fig.~\ref{fig:code_theory_overview} but in a different way. Note that we only show the models used in this paper, not the full theory space available to each code.}\label{tab:code_theory_overview}
\end{table*}

\subsection{EFTCAMB}
\texttt{EFTCAMB} is an implementation \cite{Hu:2013twa,Raveri:2014cka} of the EFT of dark energy into the CAMB \cite{2000ApJ...538..473L} EB solver (coded in fortran90) which evolves the full set of perturbations (in the synchronous gauge) arising from the action in Eq.~(\ref{eq:actionEFT}), after a built in  module checks for the stability of the model under consideration. The latter includes conditions for the avoidance of ghost and gradient instabilities (both on the scalar and tensor sector), well posedness of the scalar field equation of motion and prevention of exponential growth of DE perturbations. It can treat specific models (such as, Jordan-Brans-Dicke, designer-$f(R)$, Hu-Sawicki f(R),  Ho\v rava-Lifshitz gravity, Covariant Galileon and quintessence) through an appropriate choice of the EFT functions. It also accepts phenomenological choices for the time dependence of the EFT functions and of the dark energy equation of state which may not be associated to specific theories.

\texttt{EFTCAMB} has been used to place constraints on $f(R)$ gravity \cite{Hu:2016zrh}, Ho\v rava-Lifshitz \cite{Frusciante:2015maa}  and  specific dark energy models \cite{Raveri:2014cka}. It has also been used to explore the interplay between massive neutrinos and dark energy~\cite{Hu:2014sea}, the tension between the primary and weak lensing signal in CMB data \cite{Hu:2015rva} as well as the form and impact of theoretical priors ~\cite{Raveri:2017qvt,Peirone:2017lgi}. An up to date implementation can be downloaded from {\tt http://eftcamb.org/}. The JBD EFTCAMB solver is based on $EFTCAMB_{Oct15}$ version, while the others are based on the most recent $EFTCAMB_{Sep17}$ version.

\subsection{\texttt{hi\_class}}
\hiclass (Horndeski in the Cosmic Linear Anisotropy Solving System) is an implementation of the evolution equations in terms of the $\alpha_i(\tau)$ \cite{Zumalacarregui:2016pph} as a series of patches to the CLASS EB solver \cite{Lesgourgues:2011re,Blas:2011rf} (coded in C).
\hiclass solves the modified gravity equations for Horndeski's theory in the synchronous gauge (CLASS also incorporates the Newtonian gauge) starting in the radiation era, after checking conditions for the stability of the perturbations (both on the scalar and on the tensor sectors).
The \hiclass code has been used to place constraints on the $\alpha_i(\tau)$ with current CMB data \cite{Bellini:2015xja}, study relativistic effects on ultra-large scales \cite{Renk:2016olm}, forecast constraints with stage 4 clustering, lensing and CMB data \cite{Alonso:2016suf} and constraint Galileon Gravity models \cite{Renk:2017rzu}. 

The current public version of \hiclass is v1.1 \cite{Zumalacarregui:2016pph}. The only difference between this version and the first one (v1.0) is that v1.1 incorporates all the parametrizations used in this paper. This guarantees that the results provided in this paper are valid also for v1.0. Lagrangian-based models, such as JBD and Galileons, are still in a private branch of the code and they will be released in the future. The \hiclass code is available from \url{www.hiclass-code.net}.

\subsection{COOP}
Cosmology Object Oriented Package (COOP)~\cite{Huang:2012mt} 
is an  Einstein-Boltzmann code that solves cosmological perturbations
including very general deviations from the $\Lambda$CDM model in terms of the EFT of dark energy  parametrization \cite{Creminelli:2008wc,Gubitosi:2012hu,Gleyzes:2013ooa,Gleyzes:2014rba}. 

COOP  assumes minimal coupling of all matter species and solves the linear cosmological perturbation equations in Newtonian gauge, obtained from the unitary gauge ones by a time transformation $t \to t +\pi $. 
For the $\Lambda$CDM model, it solves the evolution equation of the spatial metric perturbation and the matter perturbation equations; details are given in Ref.~\cite{Huang:2012mt}. Beyond the $\Lambda$CDM model, COOP additionally evolves the scalar field perturbation $\pi$, using   
Eqs.~(109)--(112) of Ref.~\cite{Gleyzes:2014rba} and verifying the absence of ghost and gradient instability along the evolution.
Once the linear perturbations are solved, COOP computes CMB power spectra using a line-of-sight integral approach~\cite{Seljak:1996is, Hu:1997hp}. Matter power spectra are computed via a gauge transformation from the Newtonian to the CDM rest-frame synchronous gauge. COOP includes also the dynamics of the beyond Horndeski operator  and has been used to study the signature of a non-zero $\alpha_{\rm H}$ on the matter power spectrum as well as on the primary and lensing CMB signals \cite{DAmico:2016ntq}. COOP v1.1 has been used for this comparison. The code and its documentation are available at \url{www.cita.utoronto.ca/~zqhuang}.

\subsection{Jordan-Brans-Dicke solvers -- modified CAMB and DASh}
A systematic study, placing state of the art constraints on Jordan-Brans-Dicke gravity was presented in \cite{Avilez:2013dxa} using a modified version of
 CAMB and an altogether different EB Solver -- the Davis Anisotropy Shortcut Code (DASh)~\cite{Kaplinghat:2002mh}. DASh was initially written as a
 modification of CMBFAST \cite{1996ApJ...469..437S} by separating out the computation of the radiation and matter transfer functions from the computation of the line-of-sight integral. 
The code in its initial version, precomputed and stored the radiation and matter transfer functions on a grid so that any model was subsequently calculated 
fast via interpolation between the grid points, supplemented with a number of analytic estimates and fitting functions that speed up the calculation without significant loss of accuracy.
Such a speedup allowed the efficient traversal of large multi-dimensional parameter spaces with MCMC methods and made the study of models containing such a large parameter
 space possible~\cite{Bucher:2004an,Moodley:2004nz,Dunkley:2005va}.

The use of a grid and semi-analytic techniques was abandoned in later, not publicly available  versions of DASh, which returned to the traditional line-of-sight approach of other Boltzmann solvers.  It is possible to solve the evolution equations in both synchronous and Newtonian gauge and therefore 
is amenable to a robust internal validation of the evolution algorithm.
Over the last few years a number of gravitational theories, such as the Tensor-Vector-Scalar theory~\cite{Bekenstein:2004ne,Skordis:2008pq} 
and the Eddington-Born-Infeld theory~\cite{Banados:2008fj}, have been incorporated into the code and 
  has been recently used for cross-checks with CLASS in an extensive study of generalized dark matter \cite{Thomas:2016iav,Kopp:2016mhm}.

 In \cite{Avilez:2013dxa}, the authors used the 
internal consistency checks within DASh and the cross checks between DASh and a modified version of CAMB to calibrate and validate their results. We will use their 
modified CAMB code as the baseline against which to compare \texttt{EFTCAMB}, \hiclass and CLASSig.

\subsection{Jordan-Brans-Dicke solvers – CLASSig}

The dedicated Einstein-Boltzmann CLASSig \cite{Umilta:2015cta} for Jordan-Brans-Dicke (JBD) gravity was used in \cite{Umilta:2015cta,Ballardini:2016cvy}
to constrain the simplest scalar-tensor dark energy models with a monomial potential with the two Planck product releases and complementary astrophysical and cosmological data.
CLASSig is a modified version of CLASS which implements the Einstein equations for JBD gravity
at both the background and the linear perturbation levels without any use of approximations.
CLASSig adopts a redefinition of the scalar field ($\gamma \sigma^2 = \phi$) which recasts the original JBD theory
in the form of induced gravity in which $\sigma$ has a standard kinetic term. CLASSig implements linear fluctuations either
in the synchronous and in the longitudinal gauge (although only the synchronous version is maintained updated with CLASS).
The implementation and results of the evolution of linear fluctuations has been checked
against the quasi-static approximation valid for sub-Hubble scales during the matter dominated stage \cite{Umilta:2015cta,Ballardini:2016cvy}.
In its original version, the code implements as a boundary condition the consistency
between the effective gravitational strength in the Einstein equations at present and the one measured in a Cavendish-like experiment
($\gamma \sigma_0^2 = (1+8 \gamma)/(1+6 \gamma)/(8 \pi G)$, being $G=6.67 \times 10^{-8}$ cm$^3$ g$^{-1}$ s$^{-2}$ the Newton constant) by tuning the potential.
For the current comparison, we instead fix as initial condition $\gamma \sigma^2 (a=10^{-15})=1 \,, \dot \sigma (a=10^{-15})= 0$
consistently with the choice used in this paper.

\subsection{Covariant Galileon -- modified CAMB}
\label{sec:galcamb}

A modified version of CAMB to follow the cosmology of the Galileon models was developed in \cite{Barreira:2012kk}, and subsequently used in cosmological constraints in \cite{Barreira:2013jma, Barreira:2014jha}. The code structure is exactly as in default CAMB (gauge conventions, line-of-sight integration methods, etc.), but with the relevant physical quantities modified to include the effect of the scalar field. At the background level, this includes modifying the expansion rate to be that of the Galileon model: this may involve numerically solving for the background evolution, or using the analytic formulae of the so-called tracker evolution (see Sec.~\ref{sec:galileons_theory}). At the linear perturbations level, the modifications entail the addition of the Galileon contribution to the perturbed total energy-momentum tensor. More precisely, one works out the density perturbation, heat flux and anisotropic stress of the scalar field, and appropriately adds these contributions to the corresponding variables in default CAMB (due to the gauge choices in CAMB, one does not need to include the pressure perturbation; see \cite{Barreira:2012kk} for the derivation of the perturbed energy momentum tensor of the Galileon field). In addition to these modifications to the default CAMB variables, in the code one also defines two extra variables to store the evolution of the first and second derivatives of the Galileon field perturbation, which are solved for with the aid of the equation of motion of the scalar field, and enter the determination of the perturbed energy-momentum tensor. Before solving for the perturbations, the code first performs internal stability checks for the absence of ghost and Laplace instabilities, both in the scalar and tensor sectors.

We refer the reader to \cite{Barreira:2012kk} for more details about the model equations as they are used in this modified version of CAMB. While the latter is not publicly available\footnote{It will nonetheless be made available by the authors upon request.}, we will use this EB solver to compare codes for this class of models.

\subsection{f(R) gravity code -- {\tt CLASS\_EOS\_fR}}

{\tt CLASS\_EOS\_fR} implements the Equation of State approach (EoS) \cite{Battye:2012eu,Battye:2013aaa,Battye:2013ida} into the CLASS EB solver \cite{Blas:2011rf} for a designer $f(R)$ 
model. In the EoS approach, the $f(R)$ modifications to gravity are recast as an effective dark energy fluid 
at both the homogeneous and inhomogeneous (linear perturbation) level.

The degrees of freedom of the perturbed dark-sector are the  gauge-invariant overdensity and velocity fields, as 
described in detail in \cite{Battye:2015hza}. These obey a system of two coupled first-order differential equations, which involve the 
expressions of the gauge-invariant dark-sector anisotropic stress, $\Pi_{\rm de}$, and entropy perturbation, 
$\Gamma_{\rm de}$. The expansion of $\Pi_{\rm de}$ and $\Gamma_{\rm de}$ in terms of the other fluid degrees of freedom 
(including matter) constitute the equations of state at the perturbed level. They are the key quantities of the EoS 
approach.

The $f(R)$ modifications to gravity manifest themselves in the coefficients that appear in the expressions of 
$\Pi_{\rm de}$ and $\Gamma_{\rm de}$ in front of the perturbed fluid degrees of freedom, see \cite{Battye:2015hza} for the exact 
expressions. 
At the numerical level, the advantage of this procedure is that the implementation of $f(R)$ modifications to gravity 
reduces to the addition of two first-order differential equations to the chosen EB code (e.g.\ CLASS), while none of the 
other pre-existing equations of motion, for the matter degrees of freedom and gravitational potential, needs to be 
directly modified since it receives automatically the contribution of the total stress-energy tensor. 
In the code {\tt CLASS\_EOS\_fR}, the effective-dark-energy fluid perturbations are solved from a fixed initial time up to 
present - the initial time being chosen so that dark energy is negligible compared to matter and radiation.

At this stage, the code {\tt CLASS\_EOS\_fR} is operational for $f(R)$ models in both the synchronous and conformal Newtonian 
gauge. 
It shall soon be extended to other main classes of models such as Horndeski and Einstein-Aether theories.

A dedicated paper with details of the implementation and theoretical results and discussion is in preparation \cite{Battye:2017new}.\\

\subsection{Hořava-Lifshitz gravity code CLASS-LVDM}

This code was developed in order to test the model of dark matter with Lorentz violation (LV) proposed in Ref.~\cite{Blas:2012vn}.
The code is based on the CLASS code v1.7, 
and solves the Eqs.~(16)--(23) of Ref.~\cite{Audren:2014hza}.
The absence of instabilities is 
achieved by a proper choice of the parameters of LV in gravity and dark
matter.
All the calculations are performed in the synchronous gauge, 
and if needed, the results can be easily transformed into the Newtonian gauge.
Further details on the numerical procedure can be found in Ref.~\cite{Audren:2013dwa}
where a similar model was studied. 
The code is available at 
{\small \href{https://github.com/Michalychforever/CLASS\_LVDM}{\texttt{http://github.com/Michalychforever/CLASS\_LVDM}}}.

Compared to the standard CLASS code, one has to additionally specify four new parameters:
$\alpha,\beta,\lambda$ - parameters of LV in gravity in the khronometric model, described in Sec.\ref{sec:HLtheory},
and $Y$ - the parameter controlling the strength of LV in dark matter. 
For the purposes of this paper we switch off the latter by putting $Y\equiv 0$
and focus only on the gravitational part of khronometric/Ho\v rava-Lifshitz gravity.

The details of differences in the implementation 
w.r.t. \texttt{EFTCAMB} can be found in Appendix \ref{app:kh}.

\subsection{Non-local gravity -- modified CAMB and CLASS}
\def\nloc{RR}
\def\lcdm{\Lambda{\rm CDM}}

We compare two EB codes, a modified version of CAMB and a modified version of CLASS, that compute the cosmology of a specific model of non-local gravity modifying the Einstein-Hilbert action by a term $\sim m^2 R \Box^{-2} R$ (see Sec.~\ref{sec:NLG} for details).

The modified version of CAMB\footnote{This version of the CAMB code for the $\nloc$ model is not publicly available, but it will be shared by the authors upon request.} was developed by the authors of the GalCAMB code, and as a result, the strategy behind the code implementation is in all similar to that already described in Sec.~\ref{sec:galcamb} for the Galileon model. The strategy and specific equations used for modifying CLASS\footnote{The code is publicly available, see \cite{gitnonlocal} for the link.} are outlined in details in Appendix A of \cite{Dirian:2016puz} to which we refer the reader for an exhaustive account. In both cases, the equations that end up being coded are those obtained from the localized version of the theory that features two dynamical auxiliary scalar fields (see Sec.~\ref{sec:NL-theory}). Within both versions, the background evolution is obtained numerically by solving the system comprising the modified Friedmann equations together with the differential equations that govern the evolution of the additional scalar fields. Both implementations include a trial-and-error search of the free parameter $m$ of the model to yield a spatially flat Universe. At the perturbations level, one works out the perturbed energy-momentum tensor of the latter, and then appropriately adds the corresponding contribution to the relevant variables in the default CAMB code, whereas these have been directly put into the linearized Einstein equations in the CLASS version. The resulting equations depend on the perturbed auxiliary fields, as well as their time derivatives, which are solved for with the aid of the equations of motion of the scalar fields. The modified CAMB code was used in \cite{2014JCAP...09..031B} to display typical signatures in the CMB temperature power spectrum (although \cite{2014JCAP...09..031B} focuses more on aspects of nonlinear structure formation), whereas the modified CLASS one was used in various observational constraints studies \cite{Dirian:2014bma, Dirian:2016puz, Dirian:2017pwp}.

\section{Tests}
\label{sec:tests}
In this section we present the tests that we have performed to compare the codes described in the previous section. Ideally one should compare codes for a wide range of both gravitational and cosmological parameters. If one is to be thorough, this approach can be prohibitive computationally. Furthermore, that is not the way code comparisons have been undertaken in other situations. In practice one chooses a small selection of models and compares the various observables in these cases. This was the approach taken in the original EB code comparisons 
\cite{Seljak:2003th} but is also used in, for example, comparisons between N-body codes  for $\Lambda$CDM simulations  as well as modified gravity theories 
\cite{Winther:2015wla}. Therefore, we will follow this approach here: for each theory we will compare different codes for a handful of different parameters. 

A crucial feature of the comparisons undertaken in this section is that they always involve at least a comparison between a modified CAMB and a modified CLASS EB solver. This means that we are comparing codes which, at their core, are very different in architecture, language and genesis. For the majority of cases, we will use  \texttt{EFTCAMB} and \hiclass as the main representatives for either CAMB or CLASS but in one case (non-local gravity) we will compare two independent codes. Another aspect of our comparison is that at least one of the codes for each model is (or will shortly be made) publicly available.

\begin{figure*}[t!]
\includegraphics[width=0.9\textwidth]{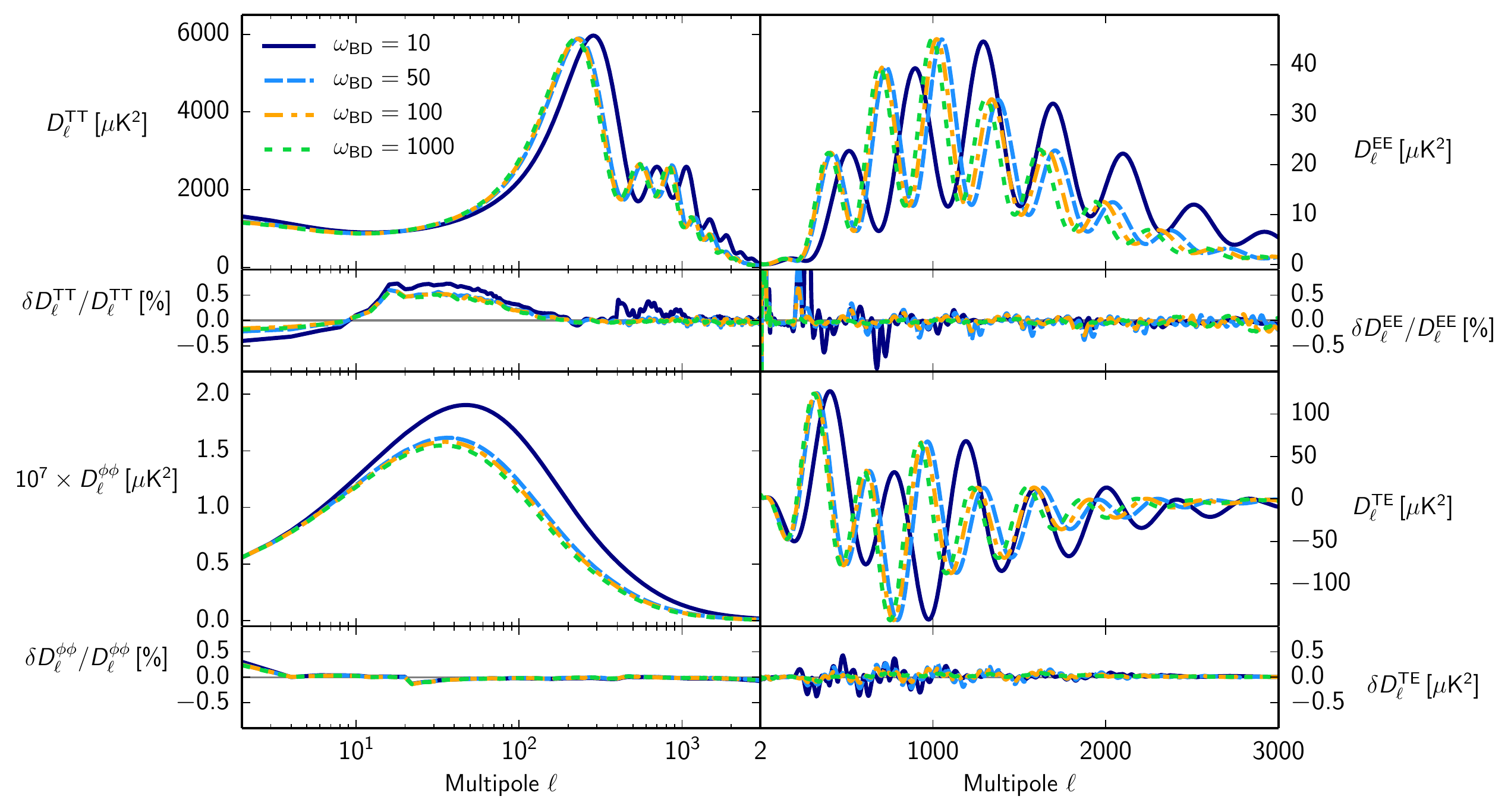}\\
\includegraphics[width=0.9\textwidth]{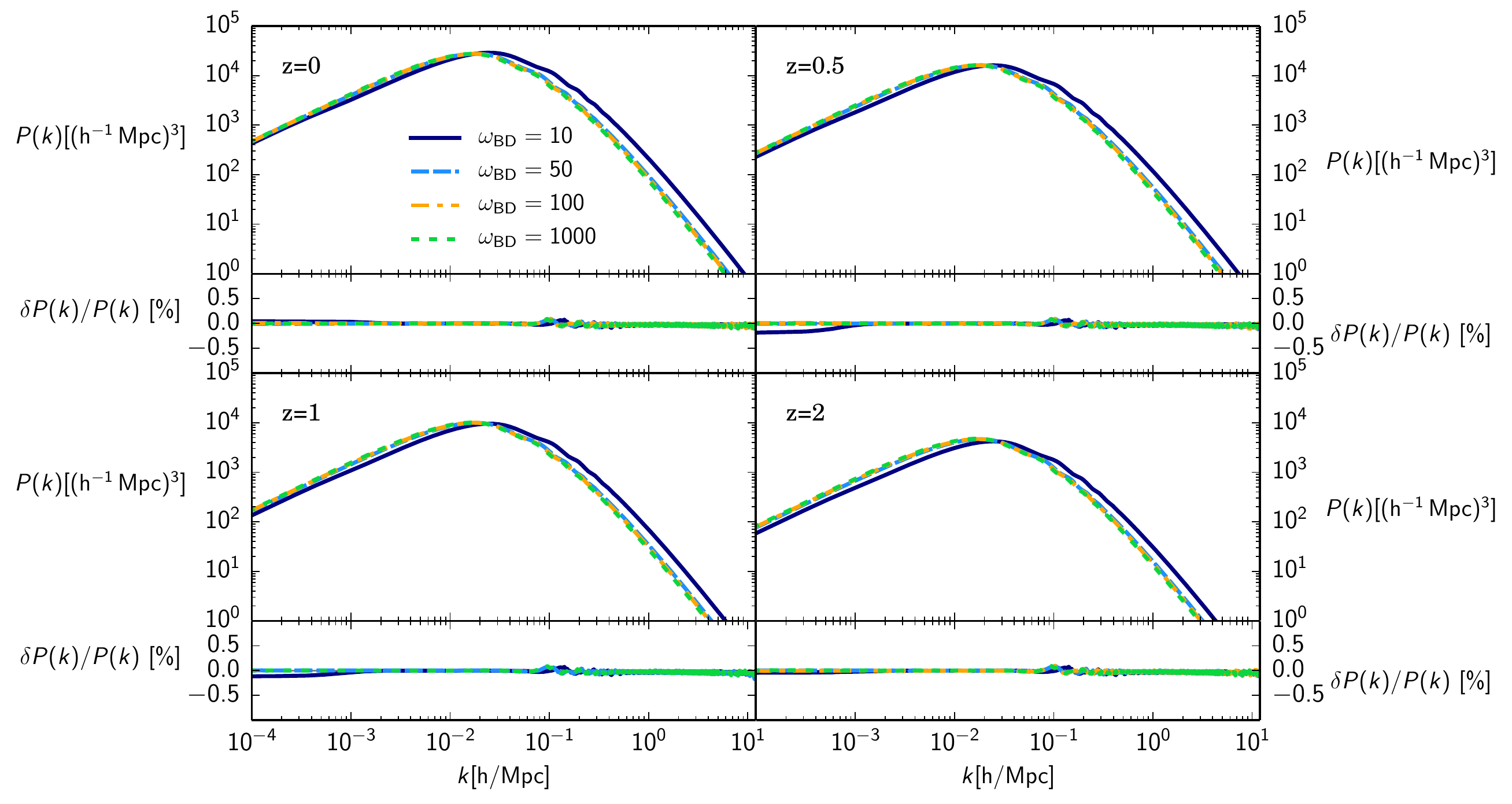}
\vspace*{-3mm}
\caption{\textbf{JBD}. Top figure: The $TT$, $EE$, lensing and $TE$ angular power spectra of the CMB -- with $D_\ell^{XY}\equiv \ell(\ell+1)/2\pi\,C_\ell^{XY}$ -- for a range of values of $\omega_{\rm BD}$ along with the relative difference between \texttt{EFTCAMB} and \hiclass. Bottom figure: The same as in the top figure but for the matter power spectrum at different redshifts. The exact values for the cosmological parameters used here can be found in Appendix \ref{sec:JBD_parameters}.}\label{fig:JBD}
\end{figure*}

\begin{figure}[ht!]
\includegraphics[width=0.8\columnwidth]{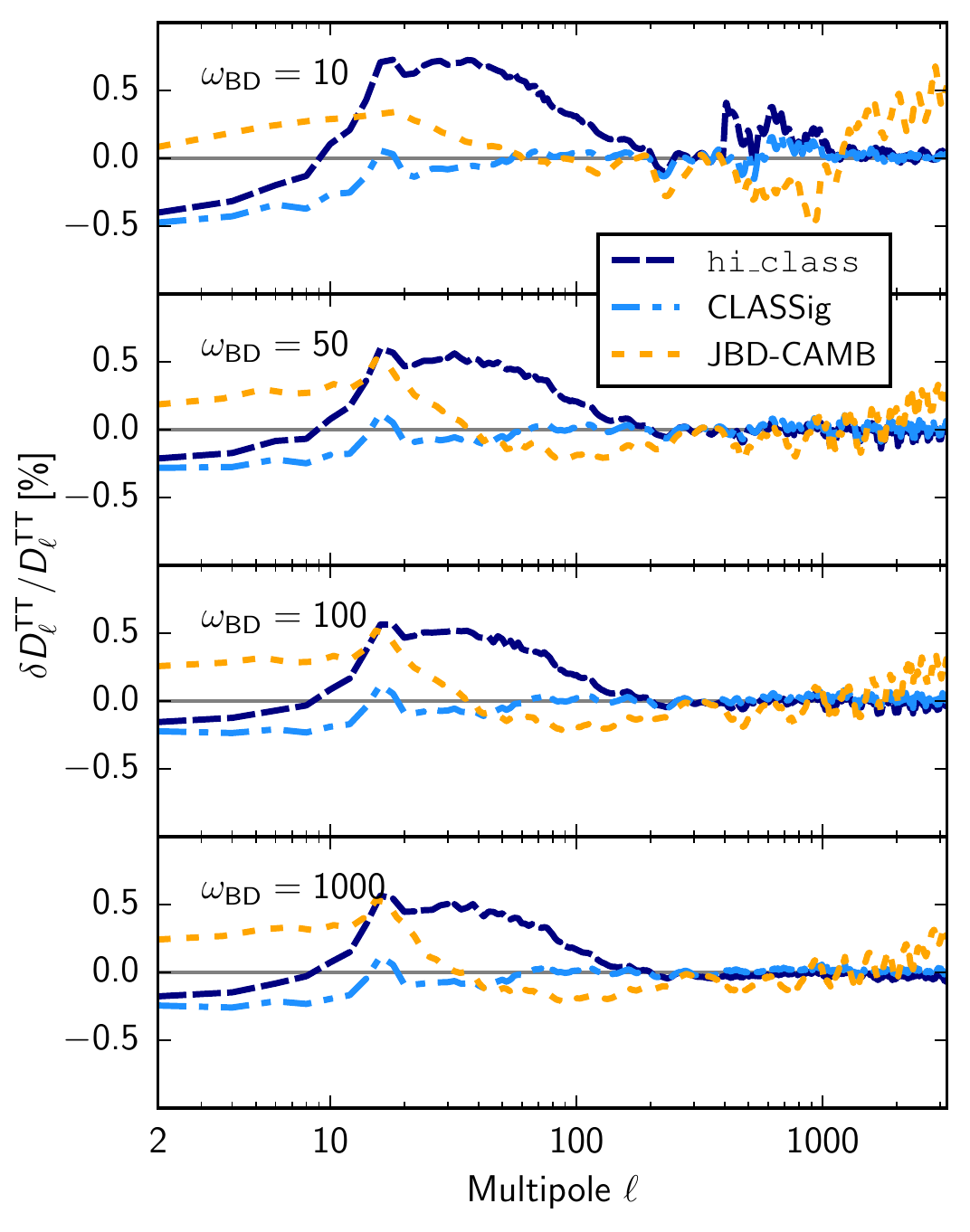}\\
\includegraphics[width=0.8\columnwidth]{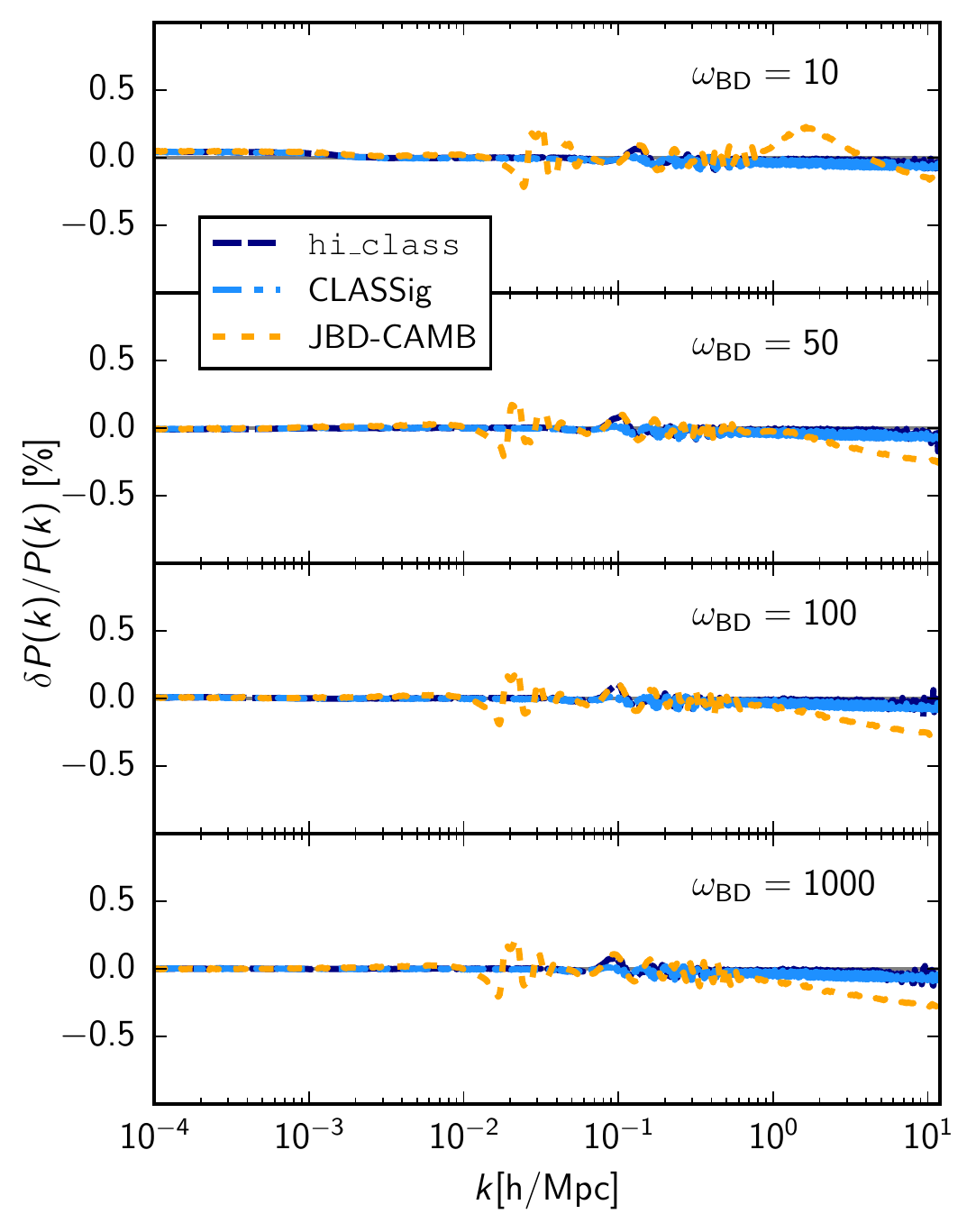}
\vspace*{-3mm}
\caption{\textbf{JBD}. Top figure: The relative difference of the $TT$ angular power spectra of the CMB for the same models showed in Fig.~\ref{fig:JBD}. Every panel corresponds to a model and the comparison of each code in the legend -- CLASSig, JBD-CAMB and \hiclass for reference -- has been done w.r.t.\ EFTCAMB. Bottom figure: The same as in the top figure but for the matter power spectrum.}\label{fig:JBD_all}
\end{figure}

In our comparisons, we will be aiming for agreement between codes -- up to $\ell=3000$ for the CMB spectra and $k=10\, h\, {\rm Mpc}^{-1}$ in the matter power spectrum  -- such that the relative distance between observables is of order $0.1\%$, with the exception of low-multipoles ($\ell<100$) where we accept differences up to $0.5\%$ since these scales are cosmic variance limited. We consider this as a good agreement, since it is smaller than the cosmic variance limit out to the smallest scales considered, i.e.~$0.1\%$ at $\ell=3000$ in the most stringent scenario (see e.g.~\cite{Seljak:2003th}). We shall see that for $\ell\lesssim300$ in the $EE$ spectra the relative difference between codes exceeds the $1\%$ bound. This clearly evades our target agreement, but it is not worrisome. Indeed, on those scales the data are noise dominated and the cosmic variance is larger than $1\%$. It is important to stress here that all the relative differences shown in the following figures are expressed in $[\%]$ units, with the exception of $\delta C_\ell^{TE}$. Since $C_\ell^{TE}$ crosses zero, we decided not to use it and to show the simple difference in $[\mu K^2]$ units instead.

Another crucial aspect has been the calibration of the codes. To do so, we fixed the precision parameters so that all the tests of the following sections (i) had at least the target agreement, and (ii) the speed of each run was still fast enough for MCMC parameter estimation. While the first condition was explained in the previous paragraph, for the latter we established a factor 3-4 as the maximum speed loss w.r.t.\ the same model run with standard precision parameters. This factor is a rough estimate that assumes that in the next years the CPU speed will increase, but even with the present computing power MCMC analysis with these calibrated codes is already possible. It is important to stress that most of the increased precision parameters are necessary only to improve the agreement in the lensing CMB spectra on small scales, which is by default 1-2 orders of magnitude worse than the other spectra.

We will be parsimonious in the presentation of results. As will become clear, we have undertaken a large number of cross-comparisons and it would be cumbersome to present countless plots (or tables). Therefore, we will limit ourselves to showing a few significant plots that help us illustrate the level of agreement we are obtaining and spell out, in the text, the battery of tests that were undertaken for each class of models. We have found our results (i.e.\ the precision with which codes agree) to be relatively insensitive to variations of the cosmological parameters.

Before showing the results of our tests, it is useful to stress here that all the precision parameters used by the codes to generate these figures are specified in Appendix~\ref{sec:precision}, while the cosmological parameters for each model are reported in Appendix~\ref{sec:parameters}.

\subsection{Jordan-Brans-Dicke gravity}\label{sec:jbd}
We have validated the \texttt{EFTCAMB}, \hiclass and CLASSig EB Solvers in two steps. We have first used DASh and the modified CAMB  of \cite{Avilez:2013dxa} to validate \texttt{EFTCAMB} with particular caveats. The current implementation of DASh uses an older version of the recombination module {\tt RECFAST} -- specifically {\tt RECFAST 1.2}. We have run \texttt{EFTCAMB} with this older recombination module and found that the agreement with DASh is at the sub-percent level. We have confirmed that this is also true in a comparison between \texttt{EFTCAMB} and the modified CAMB of \cite{Avilez:2013dxa}. We note the codes of \cite{Avilez:2013dxa} have only been cross checked and calibrated out to $\ell=2000$ and for a maximum wavenumber $k_{\rm max}=0.5\, h\, {\rm Mpc}^{-1}$. 
With the more restricted cross check of the first step in hand, we have then compared \texttt{EFTCAMB}, \hiclass and CLASSig with the more up to date recombination module  -- specifically {\tt RECFAST 1.5} -- and out to large $\ell$ and $k$. There are two main effects on the perturbation spectrum in JBD gravity: the effect of the scalar field on the background expansion and the interaction of scalar field fluctuations with the other perturbed fields. 

\begin{figure*}[t!]
\includegraphics[width=0.9\textwidth]{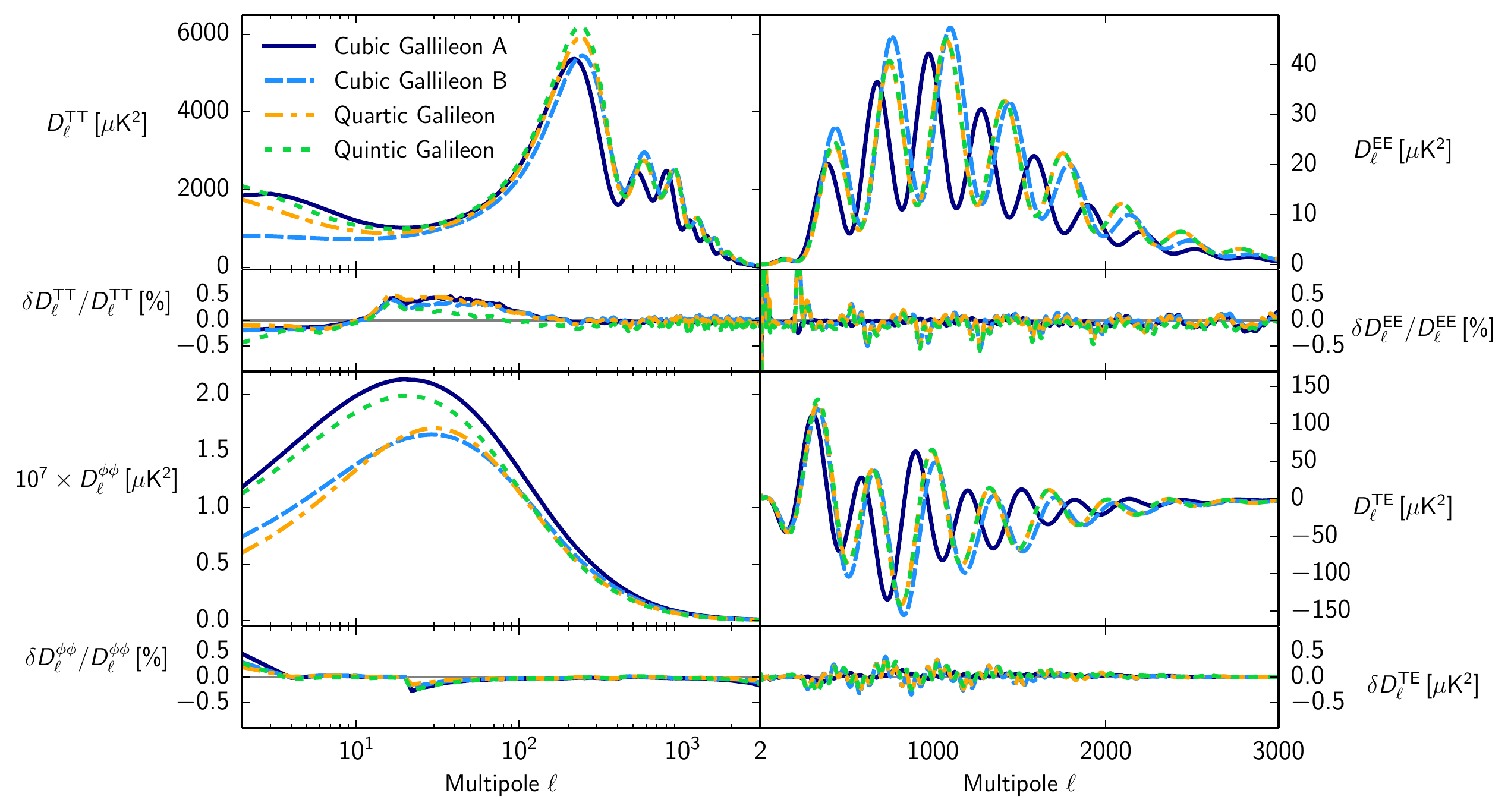}\\
\includegraphics[width=0.9\textwidth]{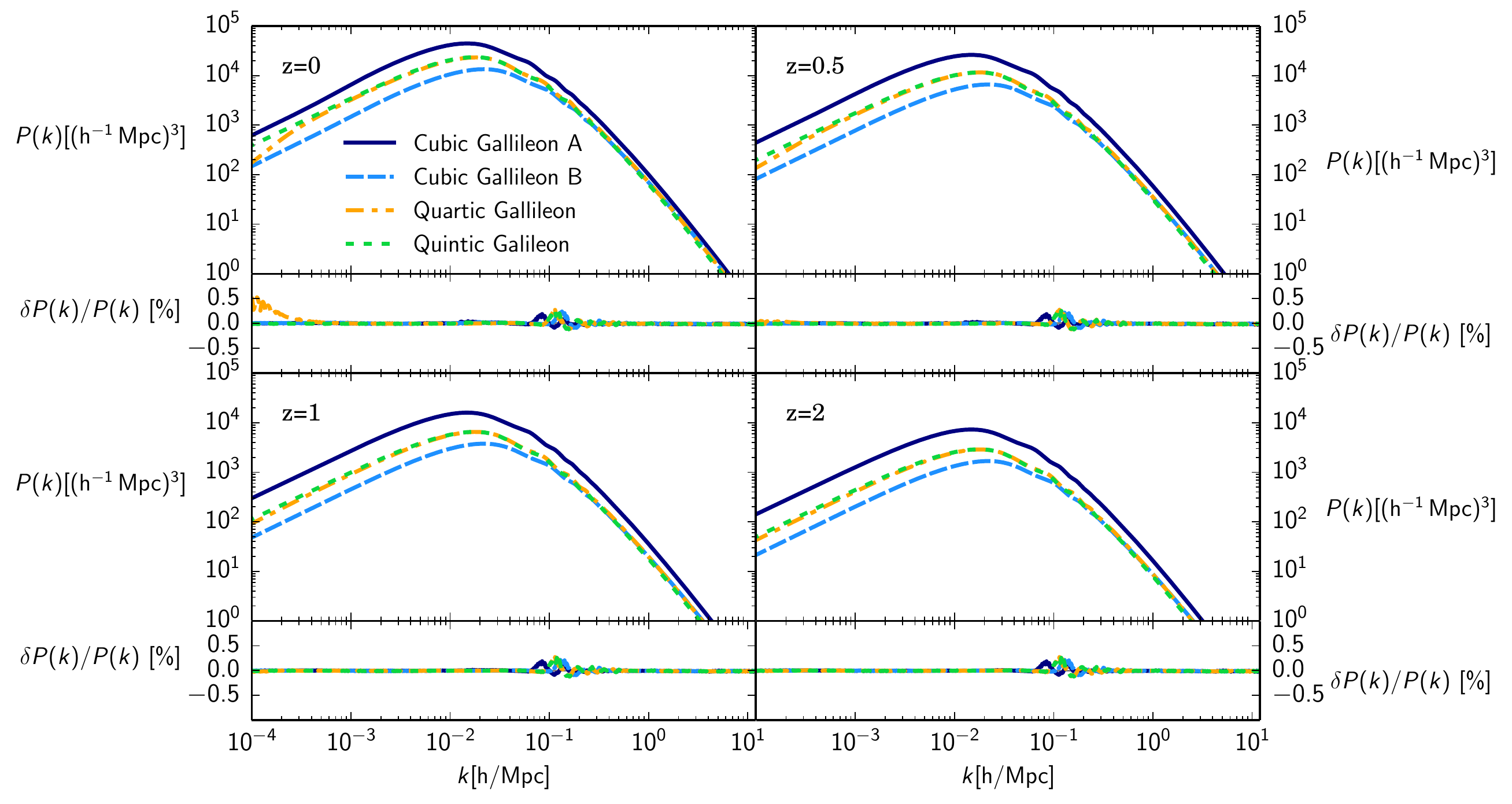}
\vspace*{-3mm}
\caption{\textbf{Covariant Galileons}. Top figure: The $TT$, $EE$, lensing and $TE$ angular power spectra of the CMB for four Galileon models along with the relative difference between \texttt{EFTCAMB} and \hiclass. Bottom figure: The same as in the top figure but for the matter power spectrum at different redshifts. The exact values for the cosmological parameters used here can be found in Appendix \ref{sec:galileon_parameters}.}\label{fig:Galileons}
\end{figure*}

\begin{figure}[ht!]
\includegraphics[width=0.8\columnwidth]{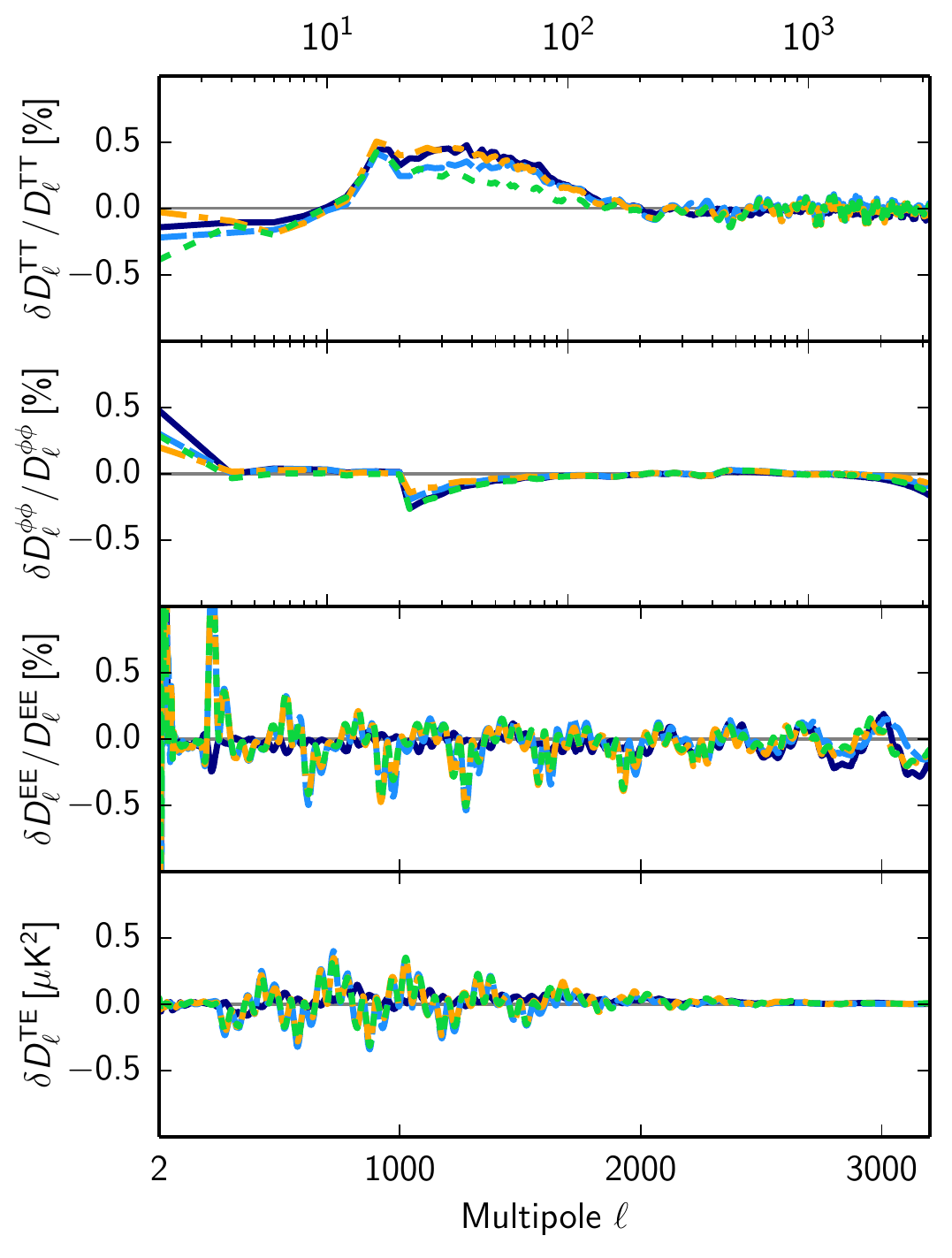}\\
\includegraphics[width=0.8\columnwidth]{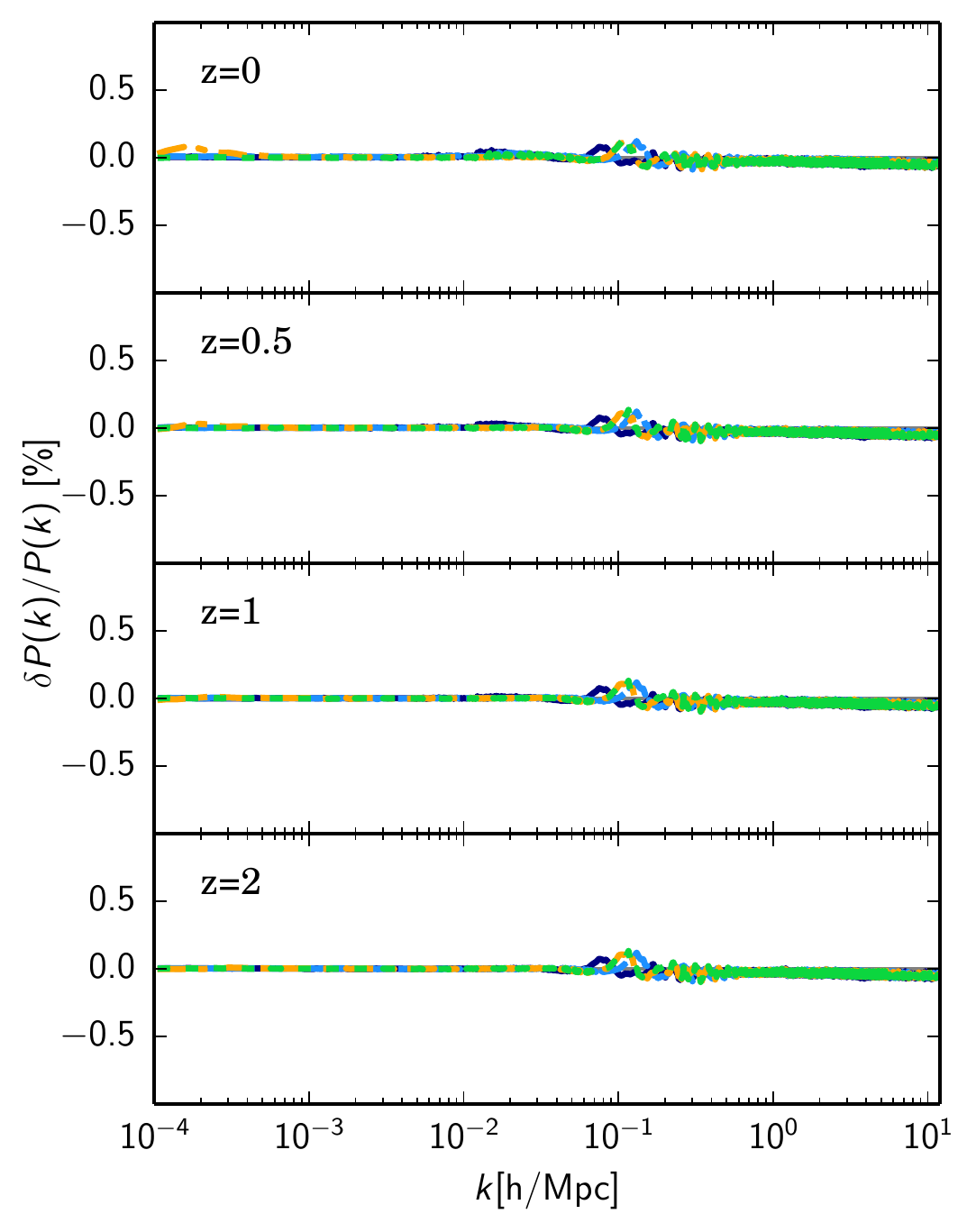}
\vspace*{-3mm}
\caption{\textbf{Covariant Galileons}. Top figure: The relative difference of the $TT$, $EE$, lensing and $TE$ angular power spectra of the CMB for the same models showed in Fig.~\ref{fig:Galileons} between GALCAMB and \hiclass (we find the same level of agreement with \texttt{EFTCAMB}). Bottom figure: The same as in the top figure but for the matter power spectrum at different redshifts.}\label{fig:Galileons_hiGal}
\end{figure}

In Fig.~\ref{fig:JBD} we show $C_\ell$ and $P(k)$  for a few different values of $\omega_{\rm BD}$ (see Appendix~\ref{sec:JBD_parameters} for the cosmological parameters used in this figures) as well as the relative difference for these quantities between \hiclass and \texttt{EFTCAMB}. We can clearly see a remarkable agreement between the codes, well within what is required for current and future precision analysis. It is possible to notice that for $\ell\lesssim10^2$ the disagreement in the temperature $C_\ell$ increases for all the models up to $\simeq0.5\%$. As we shall see, this is a common feature when comparing a CAMB-based code with a CLASS-based code, and it is present even for $\Lambda$CDM, i.e.\ using CAMB and CLASS instead of our modified versions (see e.g.\ Fig.~\ref{fig:fR}). Moreover, it has been checked that for $\Lambda$CDM a systematic bias of 1-2 orders of magnitude smaller than the cosmic variance at $\ell<100$ does not affect parameter extraction with present data, see Section 2 of \cite{Ade:2015xua}. Therefore, even if this issue deserves further investigation for DE/MG models, we believe that a better agreement at those scales is beyond the scope of this paper. The other issue of Fig.~\ref{fig:JBD}, common to all the models we show in this paper, is that the disagreement in the $C^{EE}_\ell$ on very large scales exceeds the $1\%$ bound. As we already mentioned, this is due to the fact that their amplitude approaches zero and then the relative difference is artificially boosted. This is not to worry, since (i) the amplitude of the polarization angular power spectrum is very small on large scales w.r.t.\ small scales and (ii) we are protected by cosmic variance. Finally, note that the agreement holds even for extremely small values of $\omega_{\rm JBD}$; this is essential if these codes are to be accurately incorporated into any Monte Carlo parameter estimation algorithm.

Similar results can be found in Fig.~\ref{fig:JBD_all}, where we compare the outputs of BD-CAMB, CLASSig and \hiclass (for reference) with the outputs generated by \texttt{EFTCAMB}. For simplicity, we show the result only for $C^{TT}_\ell$ and $P(k)$ at $z=0$, but the other spectra have similar behaviour as in Fig.~\ref{fig:JBD}. It is possible to note that the level of agreement is well within the $1\%$ requirement for all the codes, validating their outputs even in ``extreme'' regions of the parameter space.

This is an important first cross check between EB solvers. JBD is a canonical theory, widely studied in many regimes, and at the core of many scalar-tensor theories. It is a simple model to look at in that the background is monotonic and that only a very small subset of gravitational parameters are non-trivial.

\subsection{Covariant Galileons}\label{sec:galileons}
The Covariant Galileon theory has been implemented in the current version of \hiclass and \texttt{EFTCAMB}. Both these codes were compared against the modified CAMB described in Section \ref{sec:galcamb}, i.e.\ GALCAMB. The differences in the implementation are that \texttt{EFTCAMB} and GALCAMB assume the attractor solution for the evolution of the background scalar field, while \hiclass evolves the full background equations with the possibility of having arbitrary initial conditions. For comparison with other codes, in \hiclass we will set the initial conditions for the background scalar field as if it were on the attractor, to make the two approaches consistent and comparable. As explained in Section \ref{sec:galileons_theory}, and unlike in the JBD case (which is not self-accelerating), there is no extra parameter to vary in the case of the cubic Galileon. Once one is on the attractor and one chooses the matter densities, the evolution is completely pinned down. On the contrary, for the quartic and quintic Galileon models, there are one (for the quartic) or two (for the quintic) additional parameters. This implies that care should be had in enforcing the stability conditions (i.e.\ enforcing ghost-free backgrounds or preventing the existence of gradient instabilities).

In Fig.~\ref{fig:Galileons} it is possible to see the CMB angular power spectra and the matter power spectrum at different redshifts for two cubic Galileon models, one quartic and one quintic. While the exact values for the parameters used for this comparison are shown in Appendix~\ref{sec:galileon_parameters}, here it is important to stress that all these models have been chosen to be bad fits to current CMB and expansion history data. From these figures it can be seen that \hiclass and \texttt{EFTCAMB} agree to within the required precision. We have checked that they are also completely consistent with GALCAMB, as it is possible to see in Fig.~\ref{fig:Galileons_hiGal}, where we show the comparison between \hiclass and GALCAMB. As in the case of JBD, we have varied the cosmological and gravitational parameters and found that this agreement is robust.

\subsection{f(R) Gravity}\label{sec:fRgravity}
\begin{figure*}[t!]
\includegraphics[width=0.9\textwidth]{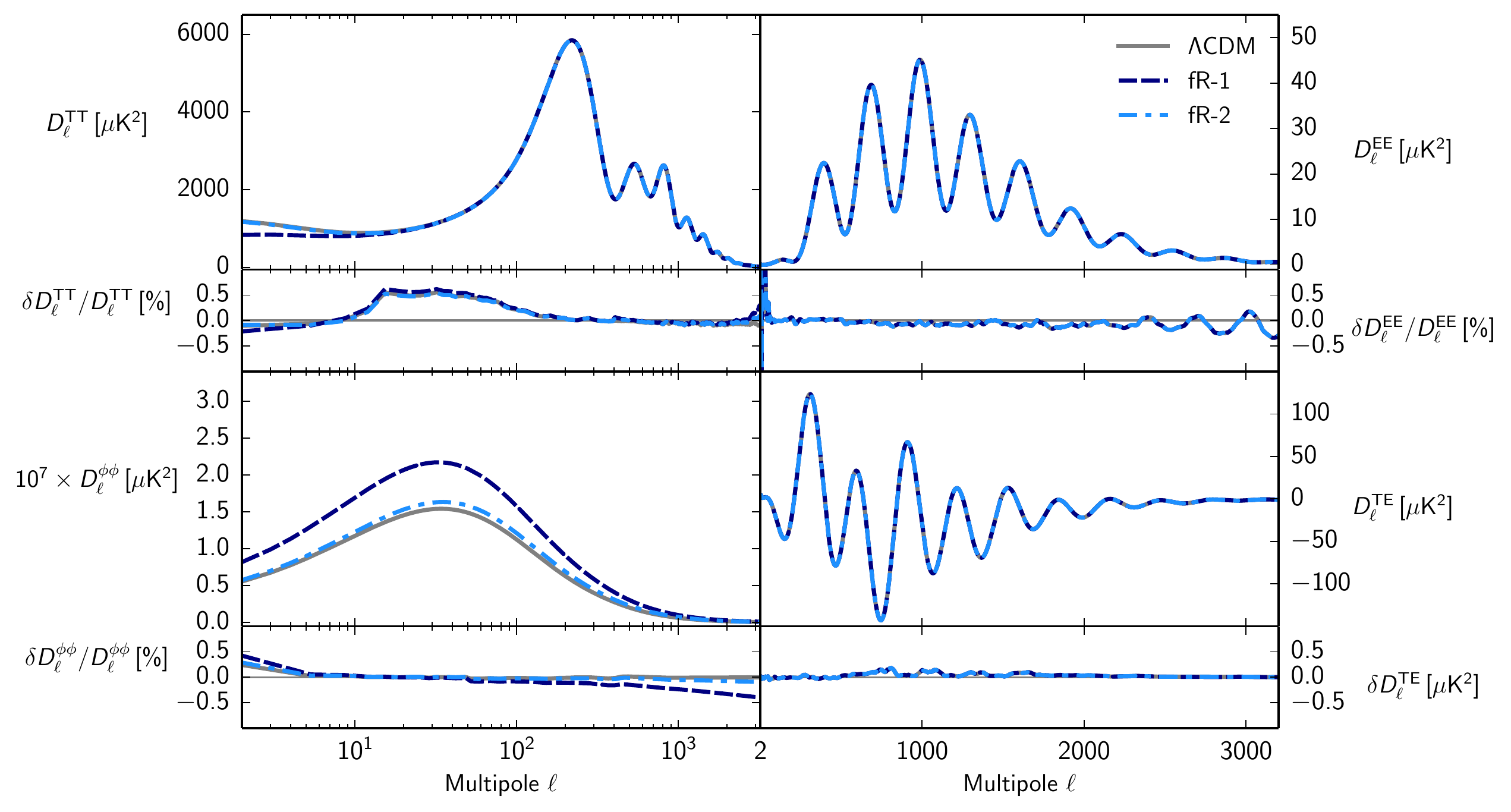}\\
\includegraphics[width=0.9\textwidth]{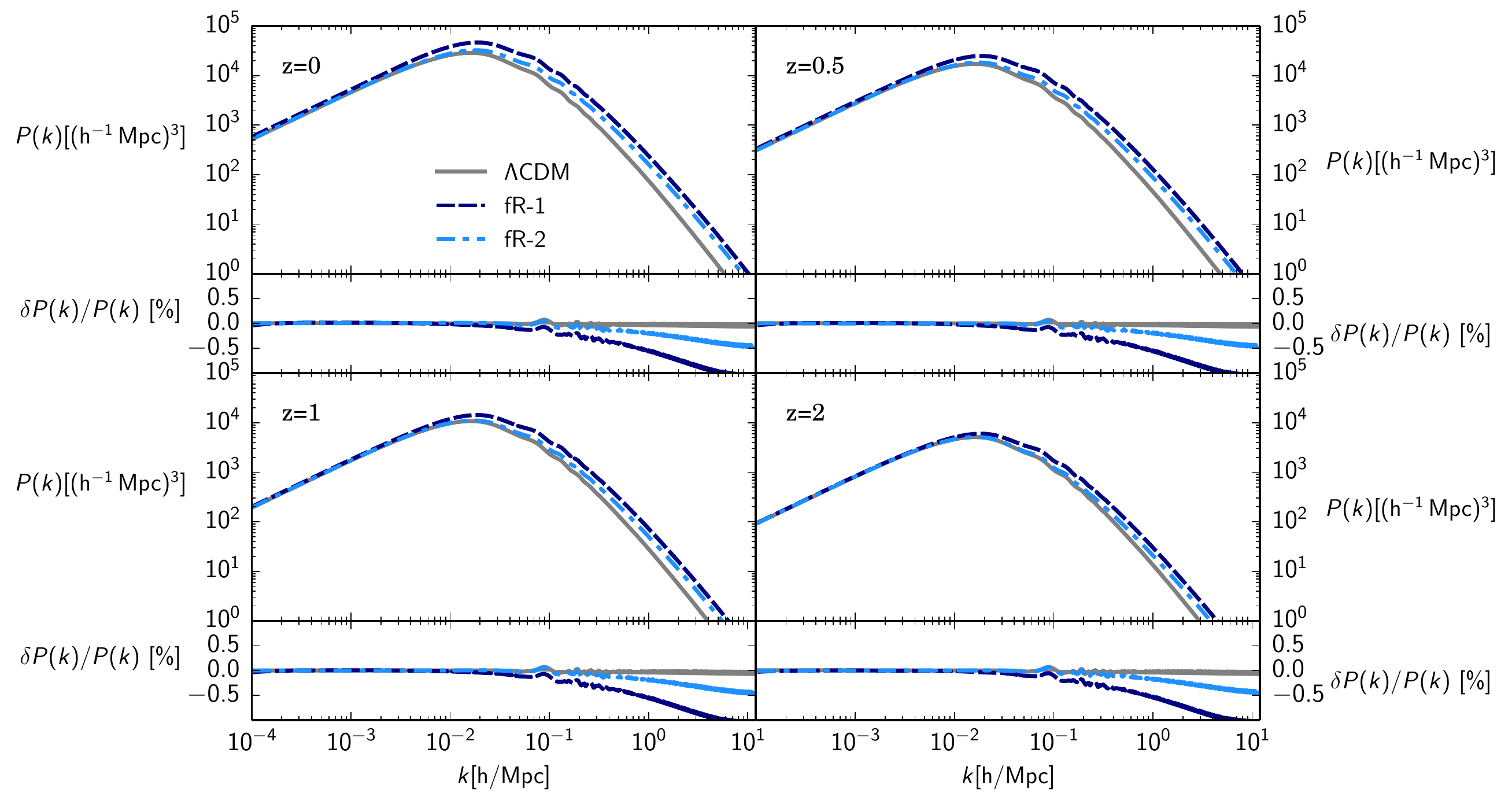}
\vspace*{-3mm}
\caption{\textbf{\textit{f(R)}}. Top figure: The $TT$, $EE$, lensing and $TE$ angular power spectra of the CMB for two different $f(R)$ models and a reference $\Lambda$CDM along with the relative difference between \texttt{EFTCAMB} and {\tt CLASS\_EOS\_fR}. Bottom figure: The same as in the top figure but for the matter power spectrum at different redshifts. The exact values for the cosmological parameters used here can be found in Appendix \ref{sec:fR_parameters}.}\label{fig:fR}
\end{figure*}

$f(R)$ gravity has been implemented in both \texttt{EFTCAMB} and {\tt CLASS\_EOS\_fR} following two independent approaches \footnote{Note that, even though $f(R)$ gravity is a sub-class of Horndeski theories, it has not been implemented in the current version of \hiclass.}.
We focus on designer $f(R)$ models that result in a $\Lambda$CDM expansion history and differ from GR at the perturbation level, displaying an enhancement of small scale structure clustering.
Once the expansion history has been chosen one has to fix a residual parameter $B_0$, corresponding to the present value of $B$, as in Eq.~(\ref{ComptonWave}).
We focus on two different values of the $B_0$ parameter: at first we compare cosmological predictions for $B_0=1$, a value that has already been excluded by experiments, to make sure no difference between the two codes is hidden by the choice of a small parameter; 
we then focus on $B_0=0.01$, that is at the boundary of CMB only experimental constraints \cite{Raveri:2014cka,Hu:2014sea} and in the range of interest for N-Body simulations.
In Fig.~\ref{fig:fR} it is possible to see that all compared spectra agree within the required precision. 
Discrepancies in all CMB spectra are consistent with the comparison to other codes and within $0.5\%$.
As in the previous cases, we have varied cosmological and gravitational parameters and found that agreement is robust. The matter power spectrum comparison shows some residual difference that reaches approximately $1\%$ on very small scales, $k=10$h/Mpc, for large values of the free parameter, $B_0=1$. The latter value is already largely excluded by CMB only data, and the scales involved are affected by non-linear clustering, hence this discrepancy is not worrisome.

\subsection{Non-local Gravity}\label{sec:NLgravity}
\label{sec:NLG}

\begin{figure*}[t!]
\includegraphics[width=0.9\textwidth]{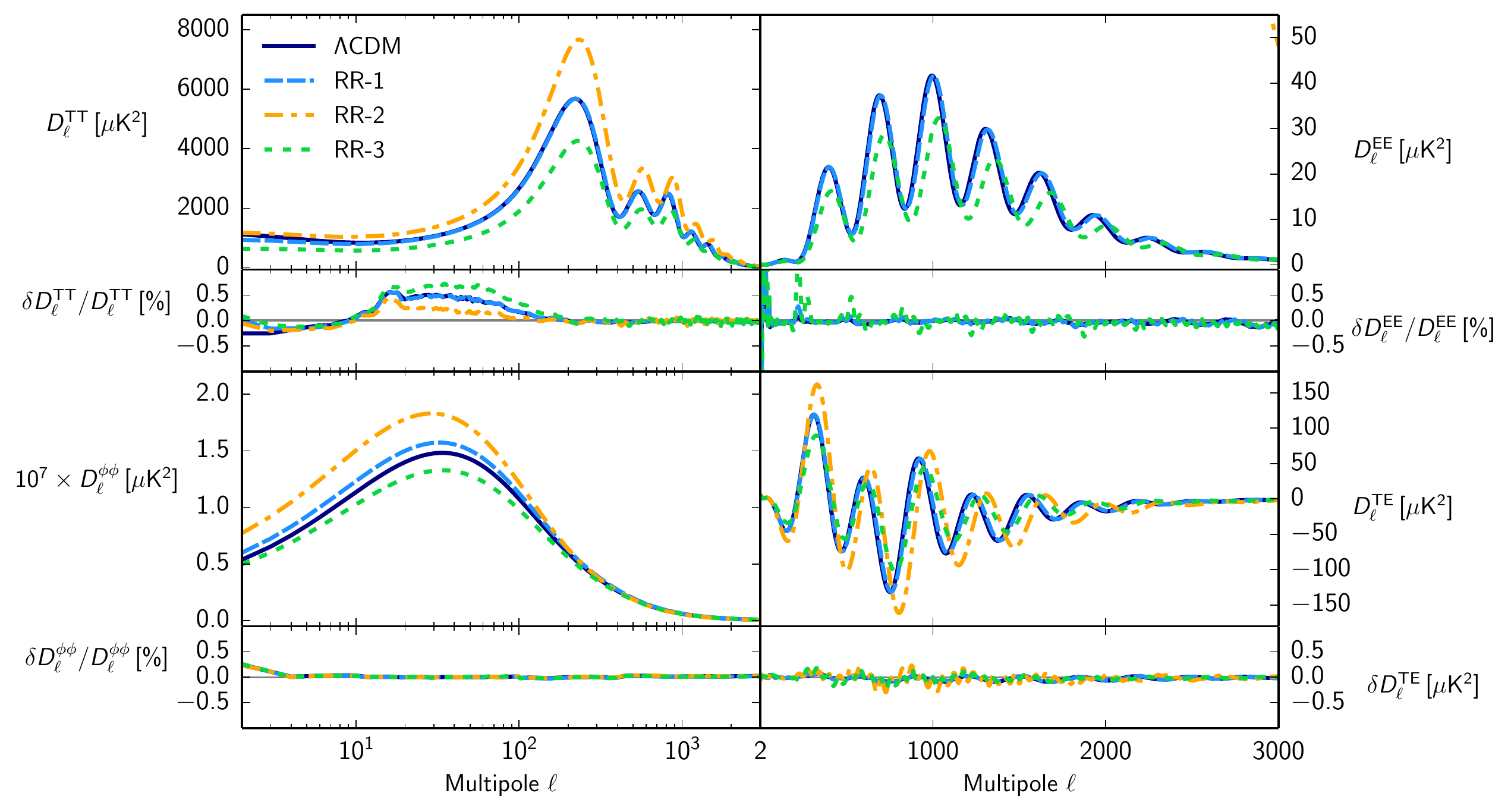}\\
\includegraphics[width=0.9\textwidth]{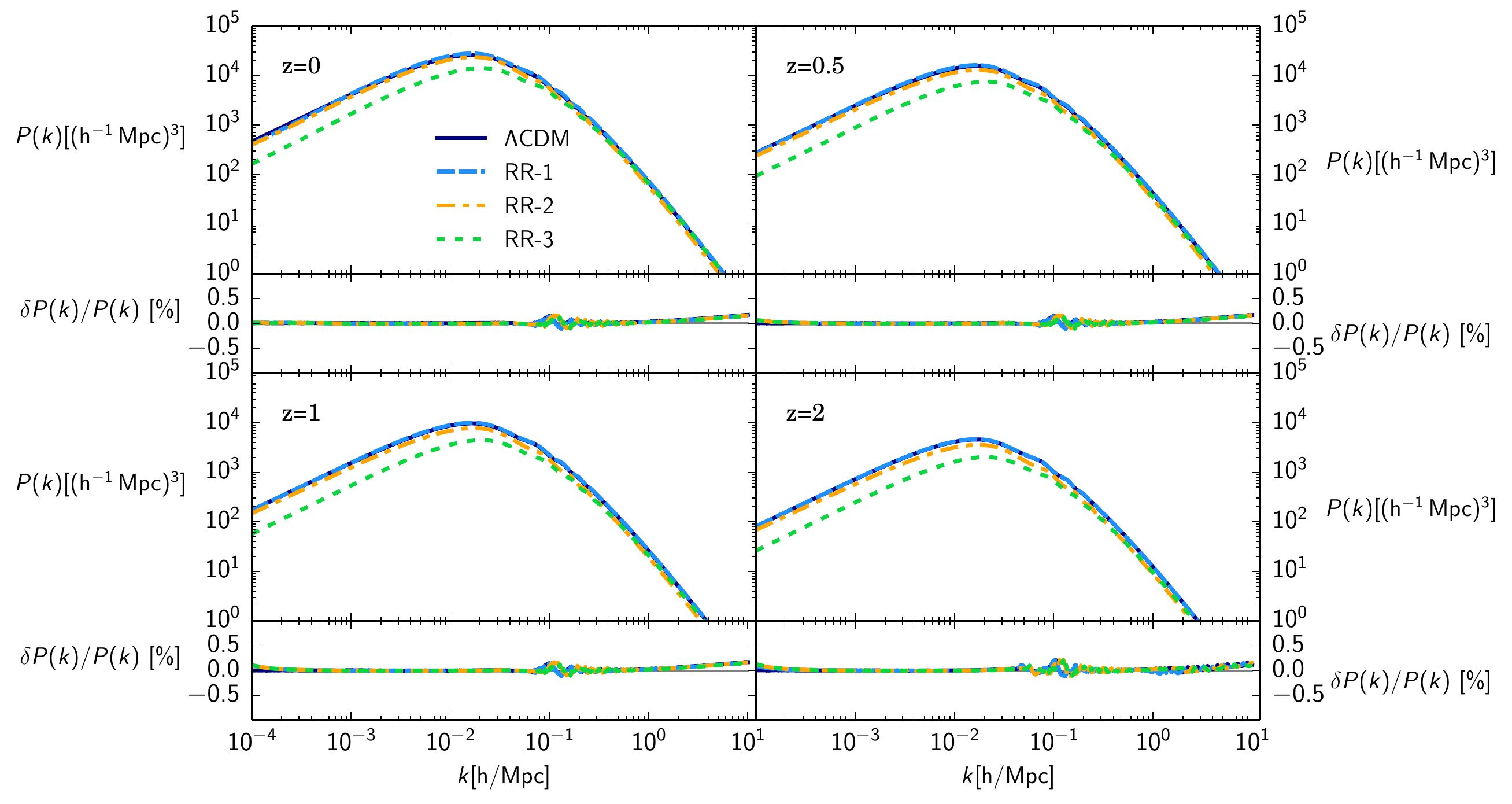}
\vspace*{-3mm}
\caption{\textbf{Non-local Gravity}. Top figure: The $TT$, $EE$, lensing and $TE$ angular power spectra of the CMB for three different Non Local Gravity models along with the relative difference between RR-CAMB and RR-CLASS. Bottom figure: The same as in the top figure but for the matter power spectrum at different redshifts. The exact values for the cosmological parameters used here can be found in Appendix \ref{sec:NL_parameters}.}\label{fig:NLG}
\end{figure*}
For the comparison of the two EB solvers of the non-local $RR$ model, we have considered three sets of cosmological parameters values, shown in Appendix~\ref{sec:NL_parameters}. Two of them are markedly poor fits to the data ($RR$-2 and $RR$-3 in Fig.~\ref{fig:NLG}, but the other gets closer to what is allowed observationally (called $RR$-1 here). In Fig.~\ref{fig:NLG}, the $\Lambda{\rm CDM}$ predictions shown correspond to the same parameters values as $RR$-1. Recall, the $\Lambda${\rm CDM} and $RR$ models have the same number of free parameters. The corresponding figures show that the level of agreement between these two EB solvers meets the required standards for all spectra, scales and redshifts shown. In fact, the shape of the relative difference curves are similar in between $\Lambda{\rm CDM}$ and the $RR$ models, which suggests that the observed differences (small as they are) are mostly due to intrinsic differences in the default codes (CAMB and CLASS), and less so due to the modifications themselves. 

\subsection{Hořava-Lifshitz Gravity}\label{sec:HLgravity}
\begin{figure*}[t!]
\includegraphics[width=0.9\textwidth]{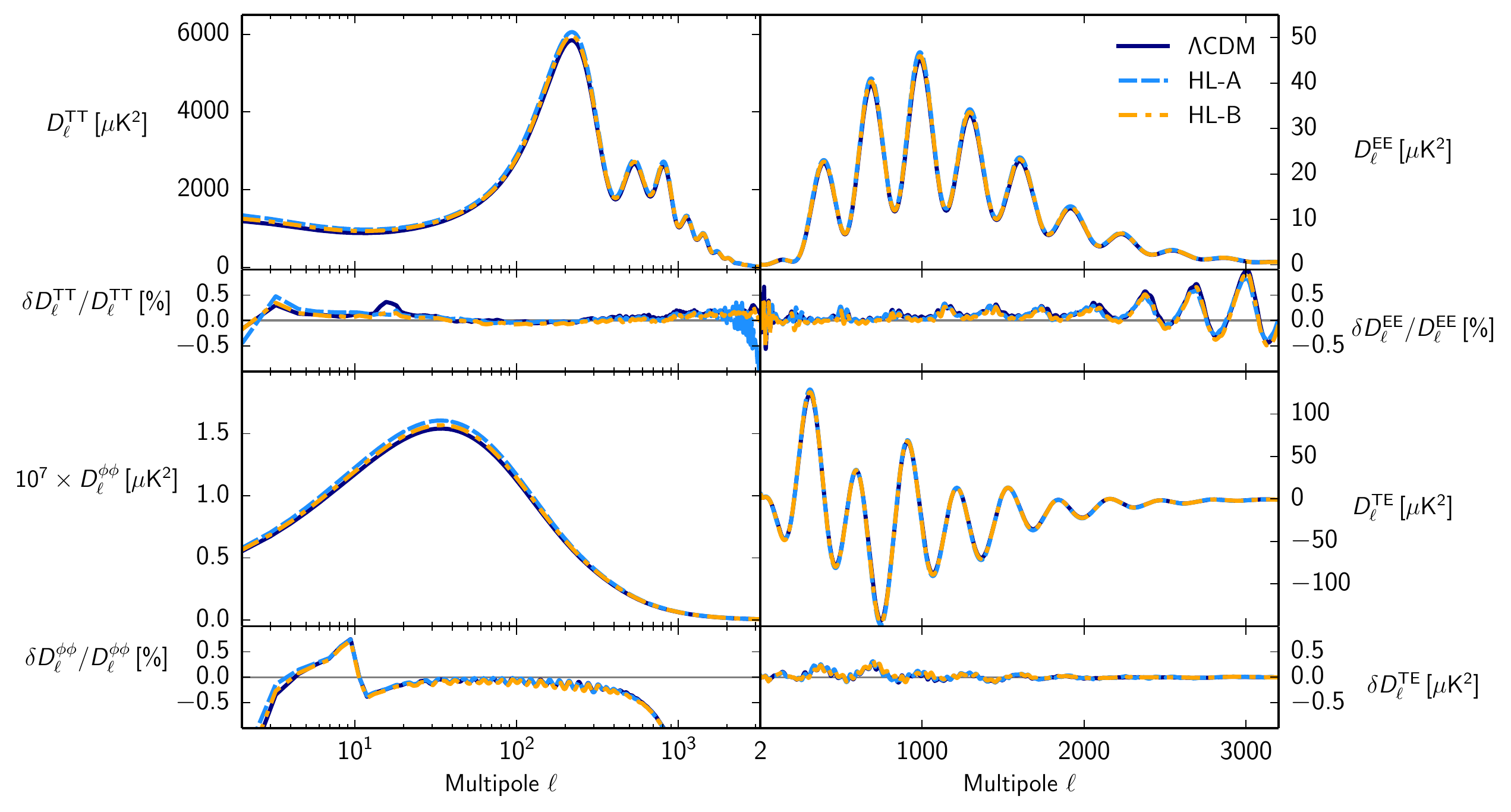}\\
\includegraphics[width=0.9\textwidth]{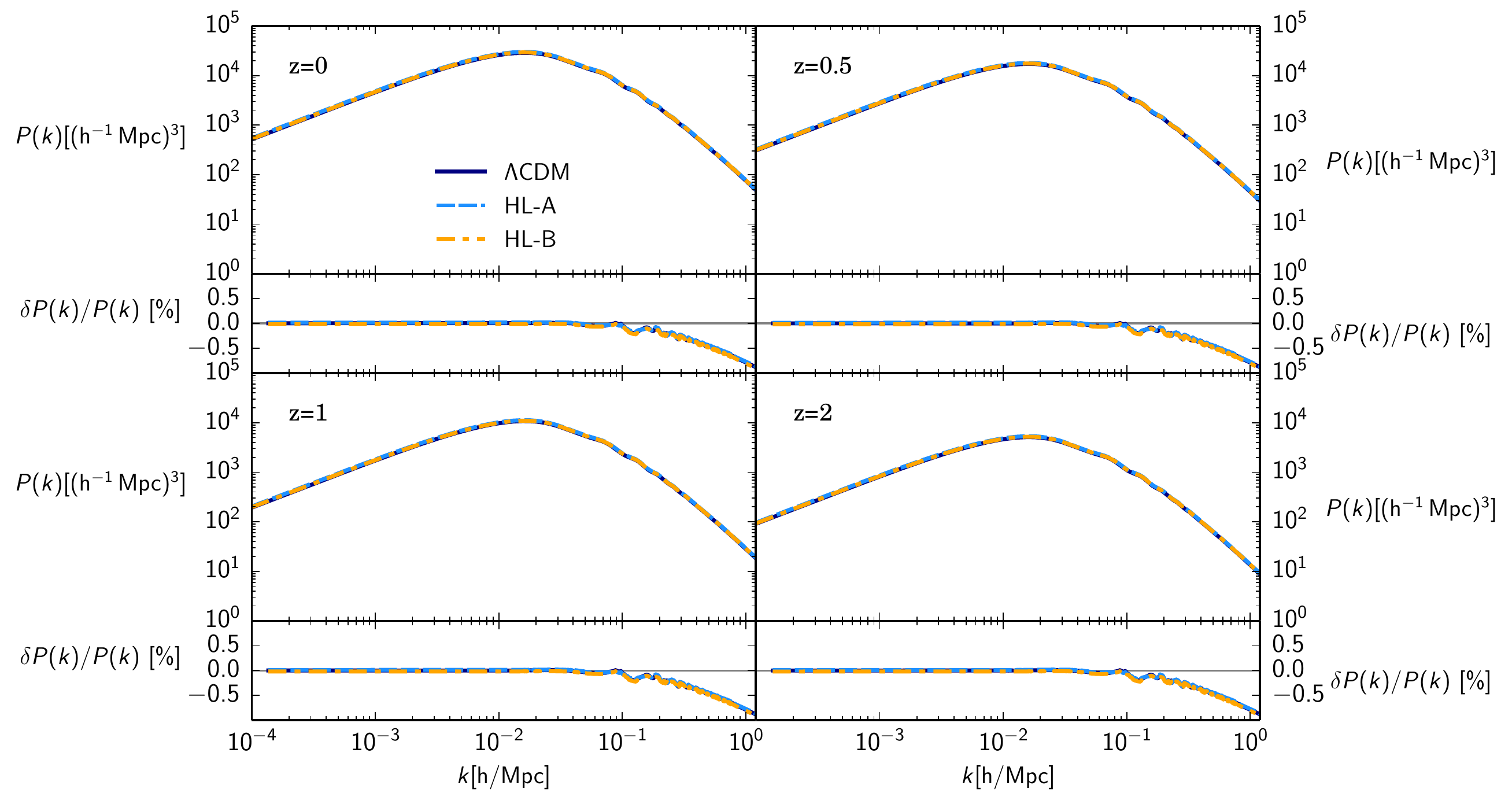}
\vspace*{-3mm}
\caption{\textbf{Ho\v{r}ava-Lifshitz Gravity}. Top figure: The $TT$, $EE$, lensing and $TE$ angular power spectra of the CMB for two different Ho\v{r}ava-Lifshitz models along with the relative difference between \texttt{EFTCAMB} and CLASS-LVDM. Bottom figure: The same as in the top figure but for the matter power spectrum at different redshifts. The exact values for the cosmological parameters used here can be found in Appendix \ref{sec:HL_parameters}.}\label{fig:HL}
\end{figure*}

We now proceed in validating EFTCAMB and CLASS-LVDM for Ho\v{r}ava-Lifshitz gravity. Because of the different implementation of the background solver (see Appendix \ref{app:kh} for details), we have limited the comparison to the subset of parameters satisfying the condition $G_{cosmo}=G_N$,  eliminating all the differences arising from it. In the top panels of Fig.~\ref{fig:HL} we compare the TT, EE, lensing and TE power spectra for two different models -- HL-A and HL-B -- and a reference $\Lambda$CDM model. These are defined by the sets of parameters specified in Appendix~\ref{sec:HL_parameters}. As we can see from the plots, the codes agree always within the $1$\% precision for TT, EE and TE power spectra. As for the lensing power spectrum we can notice an order $3$\% deviation at both small and large scales. 
Looking more carefully, one can notice that this difference is not a peculiarity of the MG model, 
but it is already present at the $\Lambda$CDM level (blue line). 
The differences at large-$\ell$ are common to all the models under investigation. 
As for the discrepancy at low-$\ell$, the fact that it is present even for $\Lambda$CDM 
suggests that it is caused by an inaccuracy in CLASS v1.7, which CLASS-LVDM is based on, and not by the modification itself.
Indeed, one may observe that this issue is absent in \hiclass based on an updated version of the CLASS code. 
In the same figure (bottom panels) we show the matter power spectra for the same models. We can see that the two codes agree well up to $k\simeq 0.1\,h\,{\rm Mpc}^{-1}$, always under the $1$\% precision. On scales $k\gtrsim 0.1\,h\,{\rm Mpc}^{-1}$ it is possible to notice that the relative differences in $P(k)$ are drastically increasing, both for $\Lambda$CDM and for the MG models. 
Like for the $C_\ell$ case, this discrepancy is due to the outdated version of the CLASS code (v1.7). For illustrative purposes we decided to cut the matter power spectrum at the value $k=1\,h\,{\rm Mpc}^{-1}$.
It should be pointed out that the scales $k\gtrsim 0.1\,h\,{\rm Mpc}^{-1}$ are significantly affected by non-linear clustering, 
therefore the output of linear Boltzmann codes in this region
is of little practical value.

Note that we used the standard CLASS accuracy flags except for lensing, where a more accurate mode has been employed 
by imposing $\texttt{accurate\_lensing = TRUE}$.

\subsection{Parametrized Horndeski functions}\label{sec:alphas}
\begin{figure*}[t!]
\includegraphics[width=0.9\textwidth]{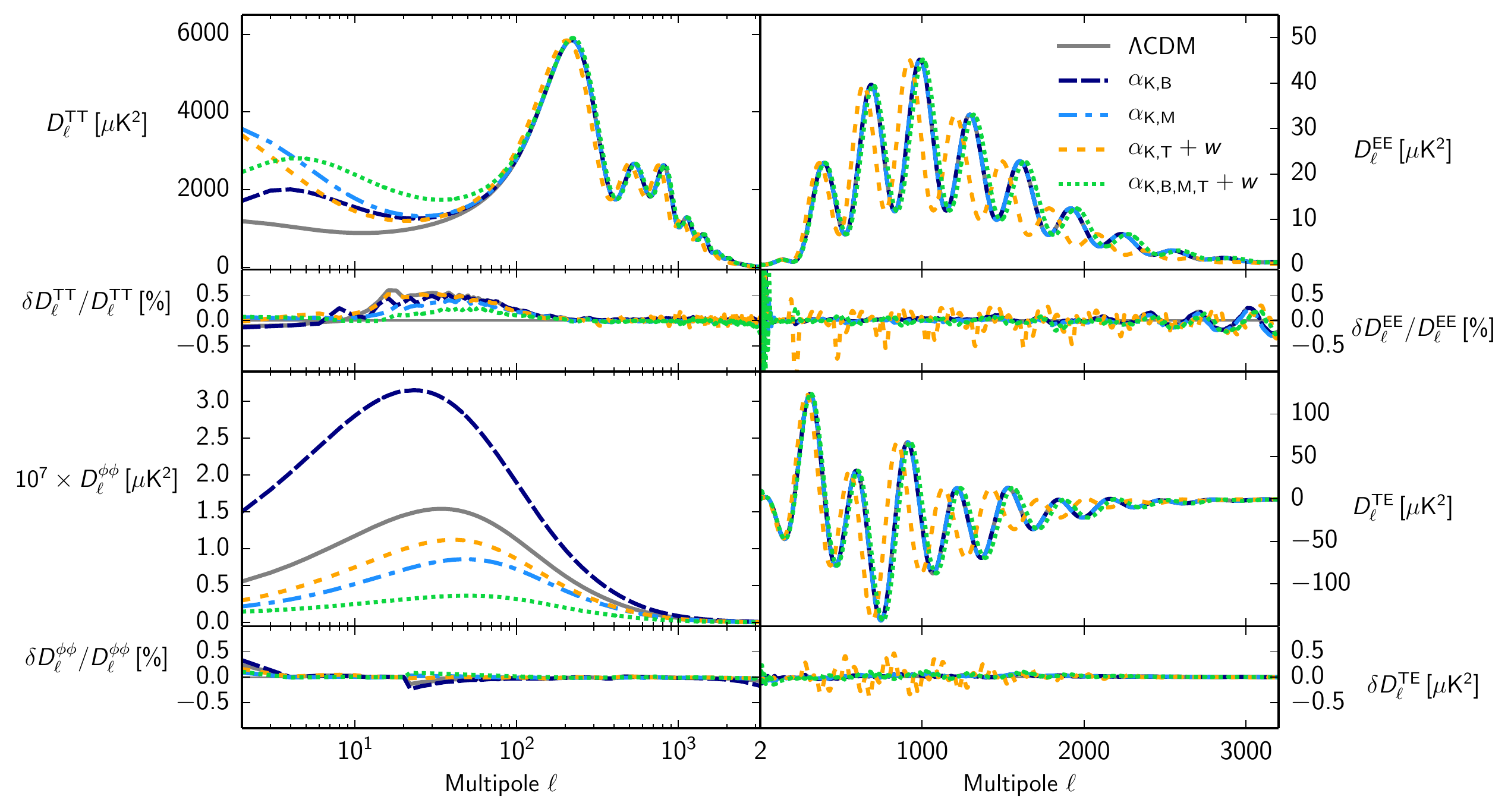}\\
\includegraphics[width=0.9\textwidth]{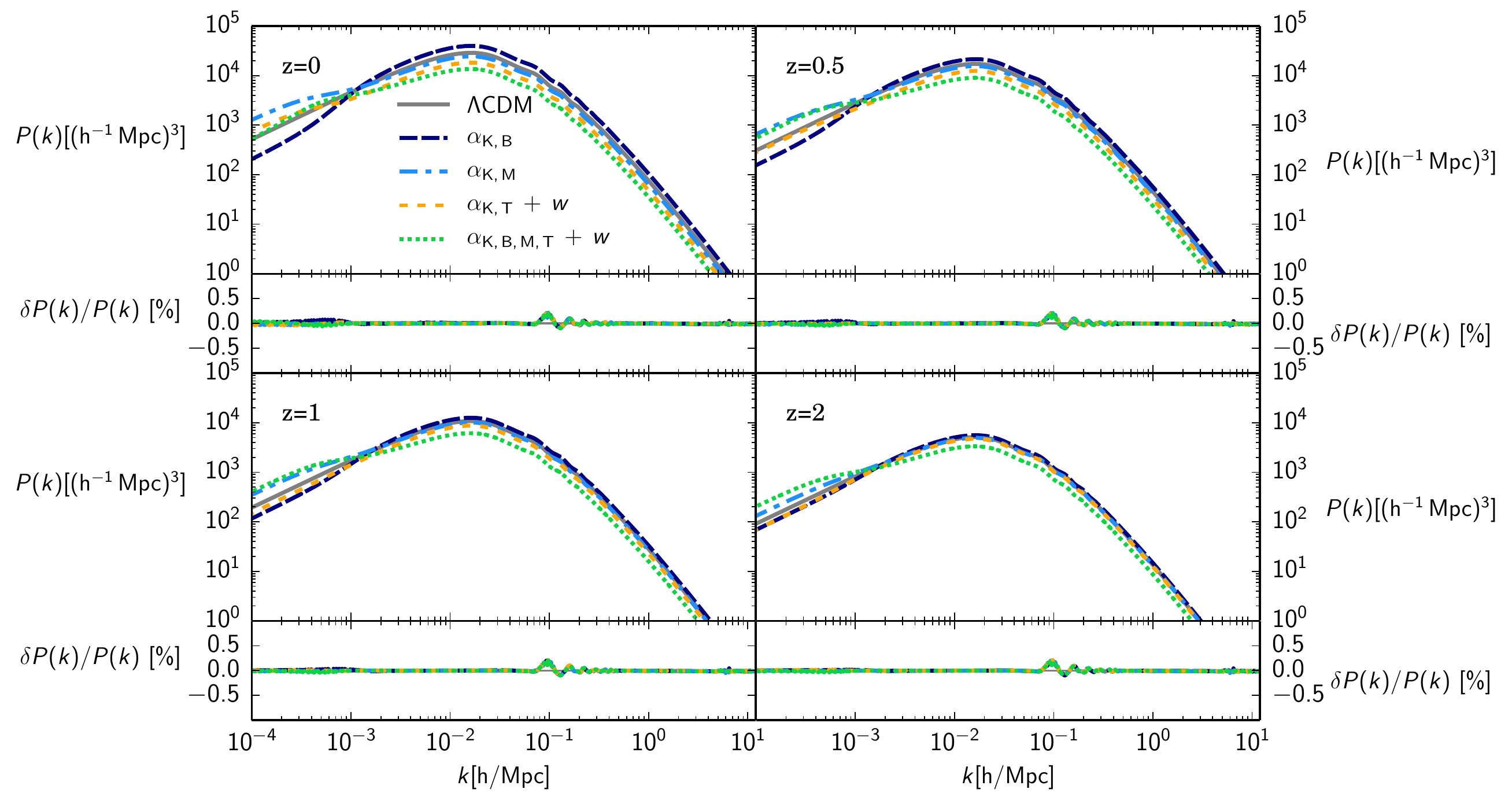}
\vspace*{-3mm}
\caption{\textbf{Alphas}. Top figure: The $TT$, $EE$, lensing and $TE$ angular power spectra of the CMB for a reference $\Lambda$CDM and four different choices of the $\{w_{\rm DE},\,\alpha_i\}$ functions along with the relative difference between \texttt{EFTCAMB} and \hiclass. Bottom figure: The same as in the top figure but for the matter power spectrum at different redshifts. The exact values for the cosmological parameters used here can be found in Appendix \ref{sec:alpha_parameters}.}\label{fig:Alphas}
\end{figure*}

\begin{figure}[ht!]
	\includegraphics[width=0.8\columnwidth]{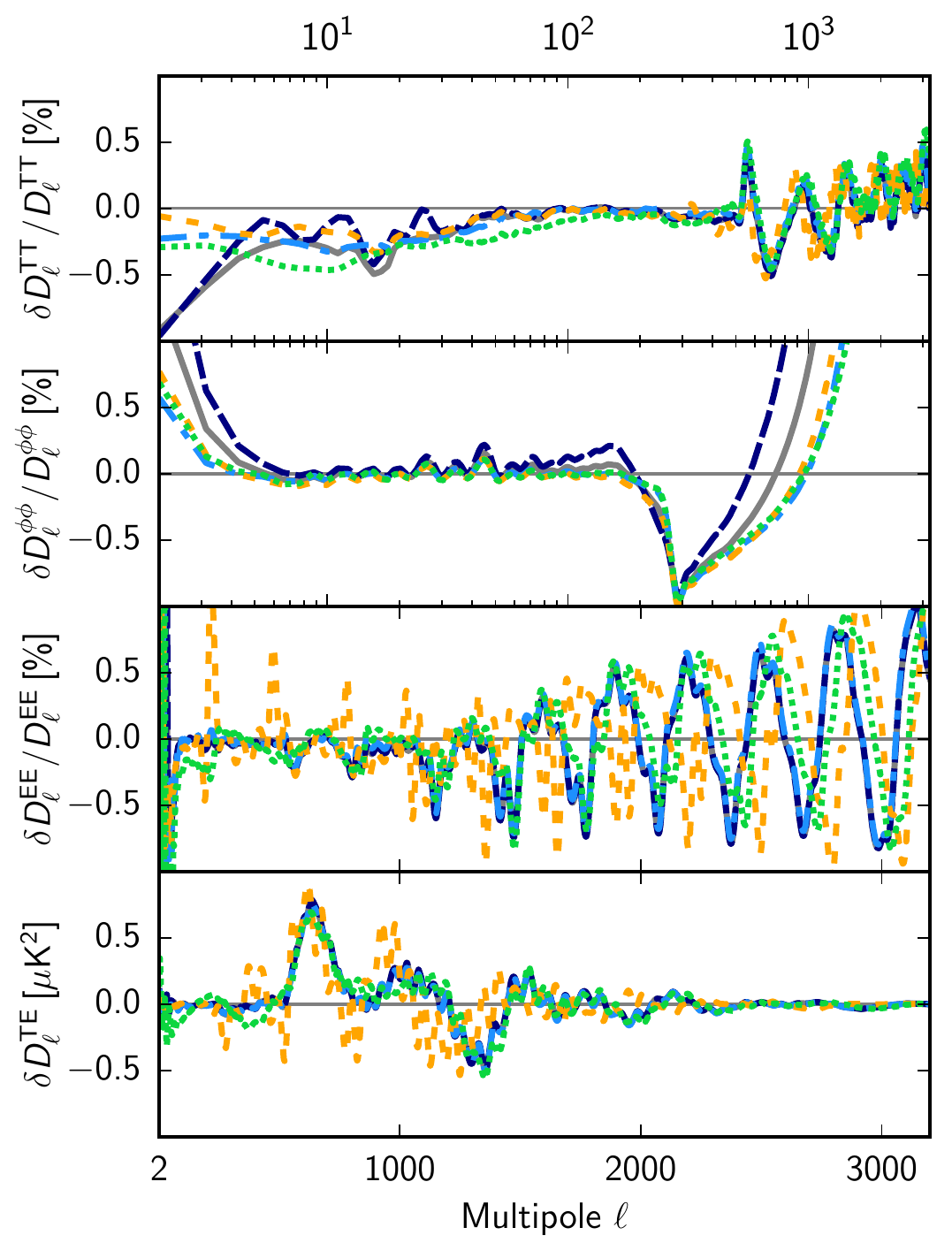}\\
	\includegraphics[width=0.8\columnwidth]{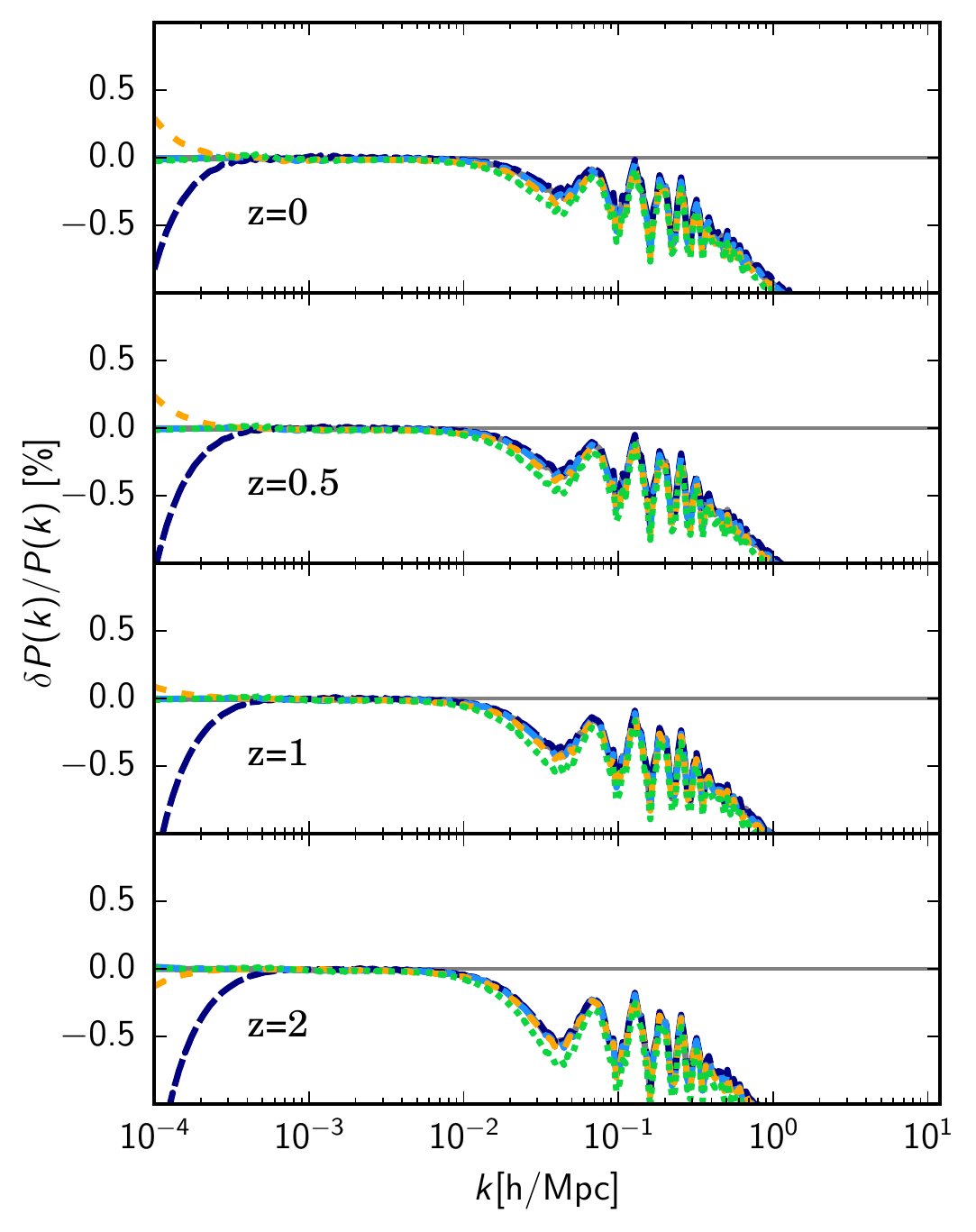}
	\vspace*{-3mm}
	\caption{\textbf{Alphas}. Top figure: The relative difference of the $TT$, $EE$, lensing and $TE$ angular power spectra of the CMB for the same models showed in Fig.~\ref{fig:Alphas} between COOP and \texttt{EFTCAMB} (we find the same level of agreement with \hiclass). Bottom figure: The same as in the top figure but for the matter power spectrum at different redshifts.}\label{fig:Alphas_eftCOOP}
\end{figure}

Up to this point we have considered a specific set of theories which, albeit representative, only involve a very restricted set of possible time evolution for either the Horndeski or EFT functions. This means that either some of the free functions are set to zero or a lower dimensional subspace of the full function space is explored (see Eq.\ (\ref{eq:JBDalphas}) for a good example). We now need to explore a wider choice of theories and time evolutions.

Ideally, we should somehow explore and compare the full parameter space described by the time dependent functions $\{\alpha_i(\tau),\,w_{\rm DE}(\tau)\}$. This is obviously impossible, but also unnecessary for our purposes. Indeed, the only modifications introduced by COOP, \texttt{EFTCAMB} and \hiclass are at the level of the Einstein and scalar field equations. Therefore, it is sufficient to use a parametrization that is capable of capturing all the terms present there. Checking that for particular parametrizations, such as rapidly varying time dependent functions, the three codes agree would in practice correspond to a  check on the differential equations solvers of each code, and this is beyond the scope of this work.

The guiding principle in choosing a particular parametrization has been to recover standard gravity at early times, to preserve the physics of the CMB and to ensure a quasi-standard evolution until recent times, i.e. approximately until the onset of dark energy. For example a parametrization closely related with this principle, which has been used in both data analysis \cite{Bellini:2015xja} and forecasts \cite{Alonso:2016suf}, takes the form
\begin{align}
w_{\rm DE} &= w_0 + (1-a) w_a \nonumber\\
\alpha_i &= c_i\Omega_{\rm DE}\,.
\end{align}
Even if this parametrization is capable of turning on all the possible freedom of Horndeski theories up to linear level, it may be not sufficient. Indeed, the system of equations for the evolution of the perturbations contains both $\{\alpha_i(\tau),\,w_{\rm DE}(\tau)\}$ and their time derivatives. Thus, we have extended this parametrization to be able to modulate the magnitude of the derivatives of these functions. The simplest choice is then
\begin{align}\label{eq:alpha_par}
w_{\rm DE} &= w_0 + (1-a) w_a \nonumber\\
M^2_*&=1+\delta M^2_0 a^{\eta_0} \nonumber \\
\alpha_i&=\alpha^0_ia^{\eta_i}\,,
\end{align}
where $i$ stands for ${K,\, B,\, T}$. The translation from the $\alpha_i$ functions to the EFT functions is provided in Appendix \ref{sec:dictionary}.

In Fig.~\ref{fig:Alphas}, we show the lensed temperature $C_\ell$ and the matter power spectrum $P(k)$ calculated at different redshifts for few different values of $\{w_0,\,w_a\}$, $\delta M^2_0$, $\alpha^0_i$ and $\eta_i$ (see Appendix~\ref{sec:alpha_parameters} for the list of values used in this comparison). The cosmological parameters are the same for each curve in the plots. The models shown in the figures were built so as to isolate the effect of each $\alpha_i$. Considering the fact that $\alpha_K$ and $\alpha_T$ alone are known to have a small effect on the observables, e.g.~\cite{Bloomfield:2012ff,Piazza:2013pua,Perenon:2015sla,Bellini:2015xja,Alonso:2016suf}, we have always combined them with other functions (either $\alpha_i$ or $w_{\rm DE}$). The $\alpha_{K,\,B,\,M,\,T}\,+\,w$ model (green dotted line) contains all the possible modifications that a Horndeski-like theory can produce. We should stress that the values used here were chosen specifically to have large deviations w.r.t.\ the reference $\Lambda$CDM model and w.r.t.\ each other. During the comparison process many more models were explored, both close to $\Lambda$CDM and unrealistically far from it.

An additional requirement to accept models for this comparison was that they were not sensitive to the specific initial conditions (ICs) set for the perturbations: The codes are set up to start with and evolve superhorizon adiabatic ICs, as predicted by standard inflation. Typically, in models which go back to GR quickly enough at early times, the other, isocurvature, modes decay with respect to the adiabatic mode, so it is irrelevant what the initial condition for the scalar field is, since it will reach the required adiabatic mode quickly.

However there are situations, typically when the modification of gravity does not decrease rapidly enough to the past, in which the isocurvature modes do not decay quickly enough (or even grow), and then it is very important that the correct, or at least equivalent, ICs be chosen.
	
The codes currently have different methods of setting ICs, which is irrelevant when the isocurvature modes decay rapidly enough, but can be important when they are not. We thus have to ensure that we are in a situation where the adiabatic ICs are an attractor for perturbations during radiation domination. The issue of setting the correct ICs for dark-energy perturbations is still an open problem and it will be addressed in future versions of the codes under consideration.

In all the cases we explored, except the ones sensitive to initial conditions as explained above, the results shown in Fig.~\ref{fig:Alphas} holds. The comparison between \texttt{EFTCAMB} and \hiclass shows a remarkable agreement, well below the $1\%$ level. It is possible to notice that the $\alpha_{K,\,T}\,+\,w$ and $\alpha_{K,\,B,\,M,\,T}\,+\,w$ models have relative differences slightly larger than the other models for the $EE$ and $TE$ CMB spectra. While it is difficult to identify one of the $\alpha_i$ or $w$ as the responsible for these deviations, we found that improving the precision parameters of each code solves this issue. This indicates that these two models are particularly complicated and they need increased precision parameters to reach the agreement of the other models. For this particular parametrization, a third code has been tested, i.e.\ COOP. The agreement between COOP and EFTCAMB is shown in Fig.~\ref{fig:Alphas_eftCOOP}. It can be noted that, even if the relative differences in CMB spectra remain below the $1\%$ level, they blow up in the matter power spectrum up to $2-3\%$ on small scales. This seems to be an effect of the accuracy of COOP. Indeed, while COOP is calibrated to get a good agreement on large scales, it lacks of precision for $k\gtrsim1hMpc^{-1}$.

\subsection{Parametrized EFT functions}\label{sec:gammas}
\begin{figure*}[t!]
\includegraphics[width=0.9\textwidth]{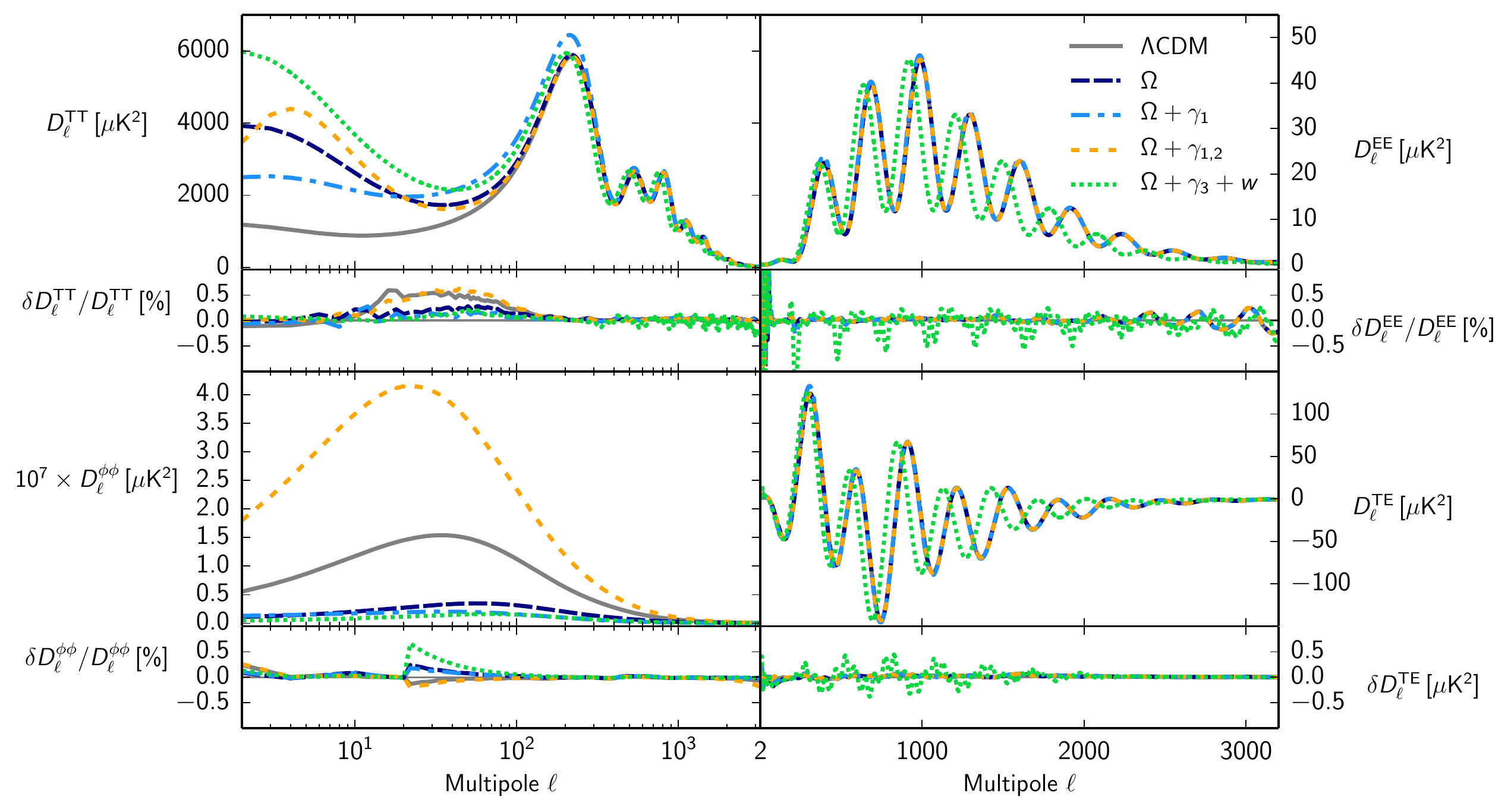}\\
\includegraphics[width=0.9\textwidth]{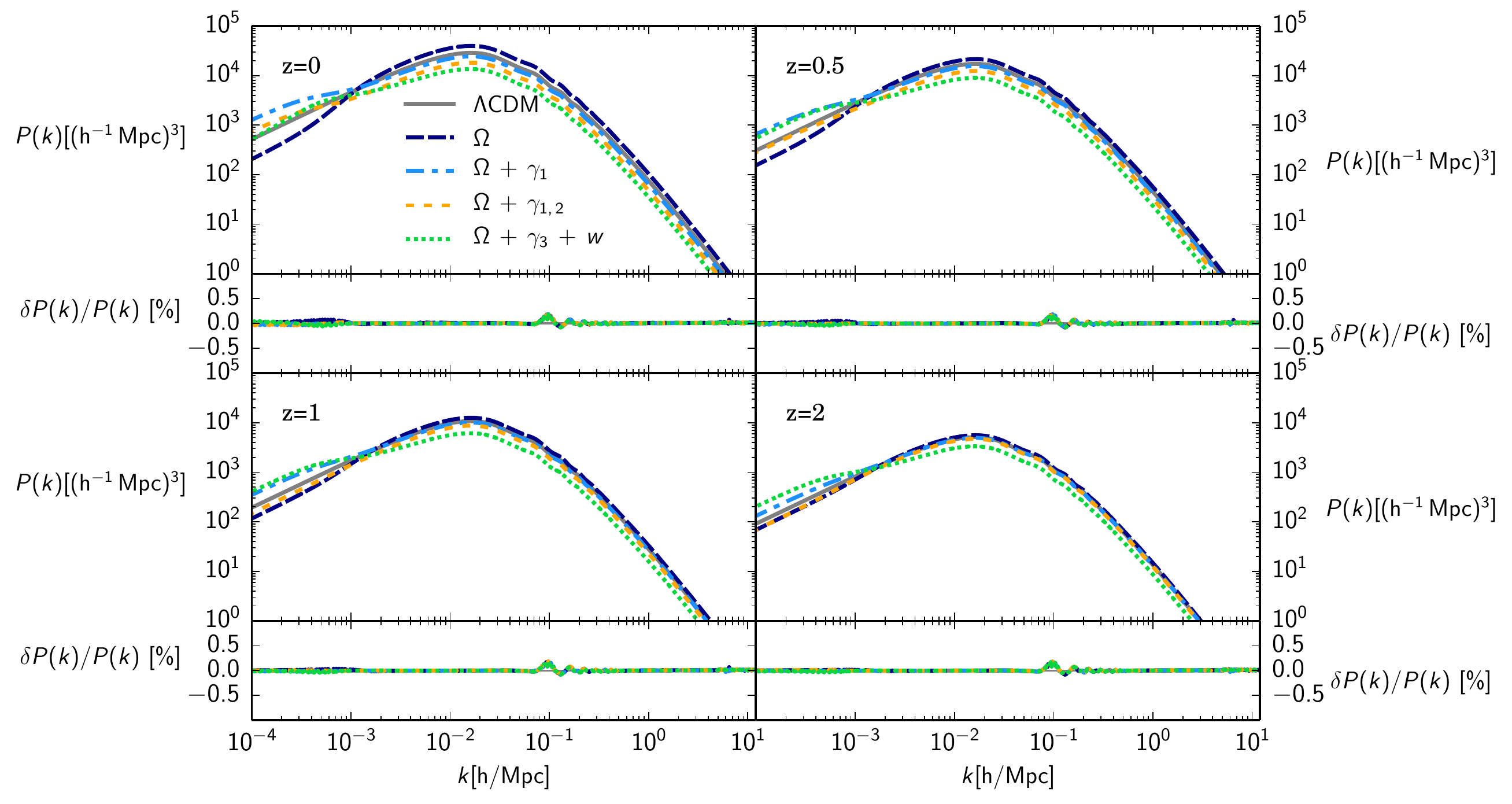}
\vspace*{-3mm}
\caption{\textbf{EFT}. Top figure: The $TT$, $EE$, lensing and $TE$ angular power spectra of the CMB for a reference $\Lambda$CDM and four different choices of the $\{w_{\rm DE},\,\Omega,\,\gamma_i\}$ functions along with the relative difference between \texttt{EFTCAMB} and \hiclass. Bottom figure: The same as in the top figure but for the matter power spectrum at different redshifts. The exact values for the cosmological parameters used here can be found in Appendix \ref{sec:gamma_parameters}.}\label{fig:Gammas}
\end{figure*}
The results presented in the previous section are able alone to establish the agreement between the three codes under consideration. However, while COOP and \hiclass were built using the $\alpha_i$ basis, \texttt{EFTCAMB} was built using the EFT approach described in Sec.~\ref{sec:eft}. As such, the structure of this code is based on $\{\Omega,\, \gamma_i\}$ functions. In case \texttt{EFTCAMB} is to be used with the $\alpha$ basis, as in the previous section, there is a built-in module which translates the $\alpha_i$ into the EFT basis before solving for the perturbations. Correspondingly \hiclass needs to  translate the $\{\Omega,\, \gamma_i\}$ functions into its preferred $\alpha_i$ basis, in order to be used for the comparison.

Let us note that when simple parametrizations are chosen, the two different bases explore different regions of the parameter space. As an example,  consider a  parametrization where $\alpha_B\propto a$. Using the conversion relations in Appendix \ref{sec:dictionary}, it is possible to show that (if $\Omega=0$) $\gamma_2\propto\mathcal{H}$, which scales as $a$ during dark-energy domination, as $a^{-1/2}$ during matter domination and as $a^{-1}$ during radiation domination.

Thus, we have also compared \texttt{EFTCAMB} and \hiclass with a particular parametrization of the $\{w_{\rm DE},\,\Omega,\,\gamma_i\}$ functions. In the same spirit as in Eqs.\ (\ref{eq:alpha_par}), we choose
\begin{align}\label{eq:gamma_par}
w_{\rm DE} &= w_0 + (1-a) w_a \nonumber\\
\Omega&=\Omega_0 a^{\beta_0} \nonumber \\
\gamma_i&=\gamma^0_ia^{\beta_i}\,,
\end{align}
where $i$ stands for ${1,\, 2,\, 3}$.

In Fig.~\ref{fig:Gammas}, we show the $TT$, $EE$, $TE$, lensing $C_\ell$'s and the matter power spectrum $P(k)$ calculated at different redshifts for a selection of different values of $\{w_0,\,w_a\}$, $\Omega_0$, $\gamma_i^0$ and $\beta_i$. The exact parameters used in these figures are shown in Appendix~\ref{sec:gamma_parameters}, and the cosmological parameters used to obtain all the curves are the same. On top of a $\Lambda$CDM reference model, the model $\Omega$ (dark blue line) represents the model used in the analysis of current data \cite{Ade:2015rim}. The other models were built to have an increasingly number of $\gamma_i$ functions and different imprints on the observables. Finally, the $\Omega\,+\,\gamma_{1,\,2,\,3}\,+\,w$ model (green dotted line) turns on all possible modifications at the same time. As in the previous section, this last model shows how model dependent are the precision parameters, having deviations in the EE and TE CMB spectra slightly larger than the other models. Within this parametrization, after neglecting all the models sensitive to the initial conditions as described in the previous section, the disagreement between \texttt{EFTCAMB} and \hiclass is within our target accuracy even for the ``extreme'' models shown in the figures.

\section{Discussion}
\label{sec:discussion}

In this paper we have shown that two general purpose publicly available EB solvers -- \texttt{EFTCAMB} and \hiclass -- are sufficiently accurate and reliable to be used to study a range of scalar-tensor theories. The third general purpose code -- COOP -- has the required precision for large scales, i.e.\ $k\lesssim1hMpc^{-1}$, but it needs to be calibrated to give accurate predictions on smaller scales. We have done this analysis by comparing these three codes to each other and to six other EB solvers that target specific theories -- DASh, BD-CAMB and CLASSig for JBD, GalCAMB for Galileons, {\tt CLASS\_EOS\_fR} for $f(R)$ and HL-CLASS for Ho\v{r}ava-Lifshitz. On top of that, we have shown that two EB solvers -- RR-CAMB and RR-CLASS -- agree very well when compared to each other for non-local gravity models. While the general principle behind these codes are similar, the implementation is sufficiently different that we believe this is a compelling validation of their accuracy. As such they are fit for purpose if we wish to analyse up and coming cosmological surveys. 

We have chosen the precision, or accuracy, settings on the codes being compared such that they could be used efficiently in a Monte Carlo Markov Chain (MCMC) analysis. It is possible to get {\it even better} agreement between the codes by boosting the precision settings. This would be done, of course, at a great loss of speed which might make the codes unusable for statistical analysis. We believe that the speed and accuracy we have achieved in this paper is a good, practical compromise. We want to emphasize here that the choice of the precision parameters is very model dependent. Indeed, for some particular configurations we had to increase somewhat the default precision to obtain agreement at the sub-percent level. If one uses the default precision parameters provided with each EB solver she might not get exactly the same agreement we have obtained in this paper. For the models we have considered, we have verified that the disagreement between the different codes was never worse than $1\%$, but it remains the responsibility of the user to verify that the precision parameters chosen are sufficient in order to obtain the accuracy desired.

Of course, there is always more to be done. We have compared these codes at specific points in model and parameter space and our hope is that they should be sufficiently stable that this comparison can be extrapolated to other models and parameters.  A possibility of taking what we have done a step further is to undertake parallel MCMC analysis with the codes being compared.
\footnote{MCMC parameter extraction has been performed on the same covariant Galileon models. The results found using modified CAMB \cite{Barreira:2014jha} and Planck 2013 data are fully consistent with those obtained with \hiclass using Planck 2015 \cite{Renk:2016olm}.}
This would fully explore the relevant parameter space and would strengthen the validation process we have undertaken in this paper. 
Furthermore, both \texttt{EFTCAMB} and \hiclass will inevitably be extended to theories beyond scalar-tensor \cite{Lagos:2016wyv,Lagos:2016gep}. The same level of rigour will need to be enforced once the range of model space is enlarged.

EB solvers can only tackle linear cosmological perturbations. There are attempts at venturing into the mildly non-linear regime using approximation schemes such as the halo-model,   perturbation theory and effective field theory of large scale structure. All attempts at doing so with the level of accuracy required by future data have focus on the standard model. There have been preliminary attempts at doing so for theories beyond GR but, it is fair to say, accurate calculations are still in their infancy. Additional complications that need to be considered when exploring this regime will be the effects of baryons, neutrinos and, more specifically, the effects of gravitational screening which can greatly modify the naive predictions arising from linear theory (a crude attempt at incorporating screening was proposed in \cite{Alonso:2016suf,Fasiello:2017bot}).

Finally, we want to emphasize that this paper is not meant to be a \textit{passepartout} to justify every kind of analysis with the codes presented here. They should not be used blindly, and we do not guarantee that all the models implemented in each version of the codes investigated here are free from bugs and reliable. When we introduced in Sec.~\ref{sec:codes} the publicly available codes we referred to a specific version, and our analysis only validates the accuracy of that version. On top of that one has to bear in mind that, even if we are quite confident that the system of equations (linearized Einstein plus scalar field equations) implemented in each code is bug free, these codes have been tested using a limited number of models. This implies that other built-in models may not be correctly implemented. So, if one wants to use one of the codes analyzed here has to follow the following steps:
\begin{enumerate}
	\item If the version of the code is not the same as the one studied here, check that it gives the same results as this version for the same models (unless this is guaranteed by the developers of the code);
	
	\item If the model that one wants to analyze has not been studied here, check that the map to convert the parameters of the models into the basis used by the code (e.g.\ $\alpha_i$ or $\gamma_i$) has been correctly implemented. Since the equations of motion are the same as used in this analysis, this is the most probable place where to find bugs, if any;
	
	\item Check that, for the model, adiabatic initial conditions are an attractor at superhorizon scales during radiation domination. If not, implement the correct initial conditons, to ensure that the addition of dark-energy isocurvature modes does not spoil predictions at late times;
	
	\item Check that the precision parameters used are sufficient to get the desired accuracy. This is very model dependent and can be done with an internal test. It is sufficient to improve them and check that the changes in the output are negligible;
	
	\item Check for a few models that the output is realistic. It can be useful to have some known limit in the parameter space to compare with.
\end{enumerate}

We believe that, with this comparison, we have placed the cosmological analysis of gravitational degrees of freedom on a robust footing. With the tools discussed in hand, we are confident that it will be possible to obtain reliable, precision constraints on general relativity with up and coming surveys.

\section*{Acknowledgements}
 EB and PGF are supported by ERC H2020 693024 GravityLS project, the Beecroft Trust and STFC. EC is supported by an STFC Rutherford Fellowship.
 BH is partially supported by the Chinese National Youth Thousand Talents Program, the Fundamental Research Funds for the Central Universities under the reference No. 310421107 and Beijing Normal University Grant under the reference No. 312232102.
 The work of MI was partly supported by the Swiss National Science Foundation and by the RFBR grant 17-02-01008.
 The research of NF is supported by Funda\c{c}\~{a}o para a  Ci\^{e}ncia e a Tecnologia (FCT) through national funds  (UID/FIS/04434/2013) and by FEDER through COMPETE2020  (POCI-01-0145-FEDER-007672).
 NF, SP and AS acknowledge the COST Action  (CANTATA/CA15117), supported by COST (European Cooperation in  Science and Technology). The work of IS and CS is supported by European Structural and Investment Funds and the Czech Ministry of Education, Youth and Sports (Project CoGraDS -- CZ.02.1.01/0.0/0.0/15\_003/0000437).
 The support by the ``ASI/INAF Agreement 2014-024-R.0 for the Planck LFI Activity of Phase E2'' is acknowledged by MB, FF and DP. MB, FF and DP also acknowledge financial contribution from the agreement ASI n.I/023/12/0 "Attivit\`a relative alla fase B2/C per la missione Euclid".
 MR is supported by U.S. Dept. of Energy contract DE-FG02-13ER41958.
 The work of NAL is supported by the DFG through the Transregional Research Center TRR33 The Dark Universe.
 Numerical work presented in this publication used the Baobab cluster of the University of Geneva. YD is supported by the Fonds National Suisse.
 UC has been supported within the Labex ILP (reference ANR-10-LABX-63) part of the Idex SUPER, and received financial state aid managed by the Agence Nationale de la Recherche, as part of the programme Investissements d’avenir under the reference ANR-11-IDEX-0004-02.
 SP and AS acknowledge support from The Netherlands Organization for Scientific Research (NWO/OCW), and from the D-ITP consortium, a program of the Netherlands Organisation for Scientific Research (NWO) that is funded by the Dutch Ministry of Education, Culture and Science (OCW).
 MB acknowledge the support from the South African SKA Project.
 MZ is supported by the Marie Sklodowska-Curie Global FellowshipProject NLO-CO.
 FV acknowledges financial support from “Programme National de Cosmologie and Galaxies” (PNCG) of CNRS/INSU, France and the French Agence Nationale de la Recherche under Grant ANR- 12-BS05-0002.
 FP acknowledges support from the post-doctoral STFC grant R120562 'Astrophysics and Cosmology Research within the JBCA 2017-2020'.

\appendix

\section{Relation between EFT functions and $\alpha$'s}\label{sec:dictionary}

In this Appendix we report the mapping between the EFT functions and the $\alpha$ bases for Horndeski theories:
\begin{eqnarray}
\Omega(a) &=& - 1 + \left(1 + \alpha_T \right)\frac{M_*^2}{M_{\rm Pl}^2}\,, \nonumber \\
\gamma_1(a) &=& \frac{1}{4a^2H_0^2M_{\rm Pl}^2}\left[\alpha_KM_*^2{\cal H}^2-2a^2c\right]\,,  \nonumber \\
\gamma_2(a) &=&  -\frac{{\cal H}}{aH_0}\left[\alpha_B \frac{M_*^2}{M_{\rm Pl}^2}+\Omega^\prime\right] \,,  \nonumber \\
\gamma_3(a) &=& -\alpha_T \frac{M_*^2}{M_{\rm Pl}^2} \,,  \nonumber \\
\gamma_4(a) &=& -\gamma_3\,, \nonumber\\
\gamma_5(a)&=&\frac{\gamma_3}{2} \nonumber\\
\gamma_6(a)&=&0,
\end{eqnarray}
where
\begin{align}
\frac{a^2 c(a)}{M_{\rm Pl}^2} =& \mathcal{H}\left(\mathcal{H}- \mathcal{H}^\prime\right) \left(1+\Omega +\frac{\Omega^\prime}{2} \right)\nonumber\\ &-\frac{\mathcal{H}^2}{2}(\Omega^{\prime\prime}-\Omega^{\prime}) -\frac{a^2 (\rho_m + p_m)}{2M_{\rm Pl}^2}\,,
\end{align}
$\alpha_M= \left(\ln M_*^2\right)^\prime$ and primes are derivatives w.r.t.\ $\ln a$. Note that the above $\alpha_B=-2\alpha_B^{\texttt{EFTCAMB}}$.

\section{Model parameters in Plots}\label{sec:parameters}

Here we list all the cosmological parameters used in this paper. For each theory we use the parameters name and the notation that can be found in Section \ref{sec:theories}.

\subsection{JBD} \label{sec:JBD_parameters}

In Section \ref{sec:jbd}, we kept fixed the following cosmological parameters:
\begin{itemize}
\item $\Omega_b h^2 = 0.02222$
 \item $\Omega_c h^2 = 0.11942$
 \item $A_s = 2.3\times 10^{-9}$
 \item $n_s = 0.9624$
 \item $\tau_{\rm reio} = 0.09$
\end{itemize}
and we varied\\
\begin{center}
\begin{tabular}{| C{1cm} | C{1.7cm} | C{1.7cm} | C{1.7cm} | C{1.7cm} |}
\hline
               & $\omega_{BD}=10$ & $\omega_{BD}=50$ & $\omega_{BD}=100$ & $\omega_{BD}=1000$ \\
\hline
$\omega_{BD}$  & $10$             & $50$             & $100$             & $1000$             \\
\hline
$H_0$          & $44.31$          & $61.43$          & $64.22$           & $66.90$            \\
\hline
\end{tabular}
\end{center}

\subsection{Covariant Galileons} \label{sec:galileon_parameters}

In Section \ref{sec:galileons}, for the Galileon models we varied all the cosmological parameters (a ``D'' in parenthesis indicates that we used that parameter as derived):\\
\begin{center}
\begin{tabular}{| C{1cm} | C{1.7cm} | C{1.7cm} | C{1.7cm} | C{1.7cm} |}
\hline
                  & Cubic Galileon A     & Cubic Galileon B     & Quartic Galileon     & Quintic Galileon     \\
\hline
$H_0$             & $75.55$              & $45$                 & $55$                 & $55$                 \\
\hline
$\Omega_b h^2$    & $0.02173$            & $0.01575$            & $0.02175$            & $0.02202$            \\
\hline
$\Omega_c h^2$    & $0.124$              & $0.100$              & $0.100$              & $0.100$              \\
\hline
$A_s$             & $2.05\times 10^{-9}$ & $2.16\times 10^{-9}$ & $2.16\times 10^{-9}$ & $2.09\times 10^{-9}$ \\
\hline
$n_s$             & $0.955$              & $0.980$              & $0.980$              & $0.954$              \\
\hline
$\tau_{\rm reio}$ & $0.052$              & $0.088$              & $0.088$              & $0.062$              \\
\hline
$\xi$             & $-2.11$ (D)          & $-1.60$ (D)          & $2.65$               & $1.4$                \\
\hline
$c_3$             & $0.079$ (D)          & $0.104$ (D)          & $-0.124$ (D)         & $0.2$                \\
\hline
$c_4$             & -                    & -                    & $-7.74\times 10^{-3}$ (D)& $0.125$ (D)      \\
\hline
$c_5$             & -                    & -                    & -                    & $-0.125$  (D)        \\
\hline
\end{tabular}
\end{center}

\subsection{f(R)} \label{sec:fR_parameters}

In Section \ref{sec:fRgravity} we kept fixed the standard cosmological parameters to these values
\begin{itemize}
 \item $H_0=69$
 \item $\Omega_b h^2 = 0.022032$
 \item $\Omega_c h^2 = 0.12038$
 \item $A_s = 2.3\times 10^{-9}$
 \item $n_s = 0.96$
 \item $\tau_{\rm reio} = 0.09$
\end{itemize}
while we varied the additional parameters\\
\begin{center}
\begin{tabular}{| C{1cm} | C{1.7cm} | C{1.7cm} | C{1.7cm} |}
\hline
           & $\Lambda$CDM    & fR-1     & fR-2    \\
\hline
$B_0$  & $0$        & $1$ &  $0.01$\\
\hline
\end{tabular}
\end{center}

\subsection{Non-Local Gravity} \label{sec:NL_parameters}

In Section \ref{sec:NLgravity} we varied all the cosmological parameters (a ``D'' in parenthesis indicates that we used that parameter as derived):
\begin{center}
\begin{tabular}{| C{1cm} | C{1.7cm} | C{1.7cm} | C{1.7cm} |}
\hline
                  & RR-1                 & RR-2                 & RR-3                 \\
\hline
$H_0$             & $67$                 & $55$                 & $55$                 \\
\hline
$\Omega_b h^2$    & $0.0222$             & $0.0222$             & $0.0222$             \\
\hline
$\Omega_c h^2$    & $0.118$              & $0.100$              & $0.120$              \\
\hline
$A_s$             & $2.21\times 10^{-9}$ & $2.51\times 10^{-9}$ & $1.81\times 10^{-9}$ \\
\hline
$n_s$             & $0.96$               & $0.93$               & $0.98$               \\
\hline
$\tau_{\rm reio}$ & $0.09$               & $0.06$               & $0.12$               \\
\hline
$m^2$             & $4.06\times 10^{-9}$ (D)          & $2.51\times 10^{-9}$ (D)          & $2.18\times 10^{-9}$ (D)          \\
\hline
\end{tabular}
\end{center}
The $\Lambda$CDM model has the same parameters as RR-1, but with a cosmological constant instead of the Non-Local parameter $m$.

\subsection{Hořava-Lifshitz gravity} \label{sec:HL_parameters}

In Section \ref{sec:HLgravity} we used the same standard cosmological parameters shown in Section \ref{sec:fR_parameters} and we varied the additional parameters\\
\begin{center}
\begin{tabular}{| C{1cm} | C{1.7cm} | C{1.7cm} | C{1.7cm} |}
\hline
           & $\Lambda$CDM    & HL-A     & HL-B    \\
\hline
$\lambda$  & $1$        & $1.2$ &  $ 1.02807 $\\
\hline
$\xi$      & $1$        & $1.3333$ & $1.05263 $\\
\hline
$\eta$    & $0$         & $0.0666$ & $0.0210526 $  \\
\hline
\end{tabular}
\end{center}

\subsection{Parametrized Horndeski functions} \label{sec:alpha_parameters}

In Section \ref{sec:alphas} we used the same standard cosmological parameters shown in Section \ref{sec:fR_parameters} and we varied the MG parameters (note that here $w_0=-1+\delta w_0$)
\begin{center}
\begin{tabular}{| C{1cm} | C{1.7cm} | C{1.7cm} | C{1.7cm} | C{1.7cm} |}
\hline
               & $\alpha_{K,B}$ & $\alpha_{K,M}$ & $\alpha_{K,T}+w$ & $\alpha_{K,B,M,T}+w$ \\
\hline
$\delta w_0$   & -              & -              & $0.9$            & $-0.5$               \\
\hline
$w_a$          & -              & -              & $-1.2$           & $1$                  \\
\hline
$\delta M_0^2$ & -              & $2$            & -                & $3$                  \\
\hline
$\eta_0$       & -              & $1.6$          & -                & $1$                  \\
\hline
$\alpha_K^0$   & $1$            & $1$            & $1$              & $1$                  \\
\hline
$\eta_K$       & $1$            & $1$            & $1$              & $1$                  \\
\hline
$\alpha_B^0$   & $1.8$          & -              & -                & $1.8$                \\
\hline
$\eta_B$       & $1.5$          & -              & -                & $1.5$                \\
\hline
$\alpha_T^0$   & -              & -              & $-0.9$           & $-0.6$               \\
\hline
$\eta_T$       & -              & -              & $1$              & $1$                  \\
\hline
\end{tabular}
\end{center}

\subsection{Parametrized EFT functions} \label{sec:gamma_parameters}

In Section \ref{sec:gammas} we used the same standard cosmological parameters shown in Section \ref{sec:fR_parameters} and we varied the MG parameters (note that here $w_0=-1+\delta w_0$)
\begin{center}
\begin{tabular}{| C{1cm} | C{1.7cm} | C{1.7cm} | C{1.7cm} | C{1.7cm} |}
\hline
               & $\Omega$ & $\Omega+\gamma_1$ & $\Omega+\gamma_{1,2}$ & $\Omega+\gamma_{3}+w$ \\
\hline
$\delta w_0$   & -        & -                 & -                     & $0.9$                 \\
\hline
$w_a$          & -        & -                 & -                     & $-1.2$                \\
\hline
$\Omega_0$     & $2$      & $1$               & $2$                   & $2$                   \\
\hline
$\beta_0$      & $1$      & $0.4$             & $1.5$                 & $1$                   \\
\hline
$\gamma_1^0$   & -        & $1$               & $1$                   & -                     \\
\hline
$\beta_1$      & -        & $1$               & $1$                   & -                     \\
\hline
$\gamma_2^0$   & -        & -                 & $-4.8$                & -                     \\
\hline
$\beta_2$      & -        & -                 & $0$                   & -                     \\
\hline
$\gamma_3^0$   & -        & -                 & -                     & $2$                   \\
\hline
$\beta_3$      & -        & -                 & -                     & $1$                   \\
\hline
\end{tabular}
\end{center}

\section{Precision parameters in Plots}\label{sec:precision}
In order to improve the accuracy of the results, keeping in mind that the CPU-time should remain acceptable for MCMC runs, we changed the default values for some precision parameter
\begin{itemize}
 \item CAMB-based codes
  \begin{verbatim}
get_transfer = T
transfer_high_precision = T
high_accuracy_default = T
k_eta_max_scalar = 80000
do_late_rad_truncation = F
accuracy_boost = 1
l_accuracy_boost = 1
l_sample_boost = 1
l_max_scalar = 10000
accurate_polarization = T
accurate_reionization = T
lensing_method        = 1
massive_nu_approx     = 0
use_spline_template   = T
accurate_BB           = F
EFTCAMB_turn_on_time =   1e-10
  \end{verbatim}
 \item CLASS-based codes (except for CLASS-LVDM)
  \begin{verbatim}
l_max_scalars = 5000
P_k_max_h/Mpc = 12.
perturb_sampling_stepsize = 0.010
l_logstep=1.026
l_linstep=25
l_switch_limber = 20
k_per_decade_for_pk = 200
accurate_lensing = 1
delta_l_max = 1000
k_max_tau0_over_l_max=8 
  \end{verbatim}
\end{itemize}

\section{Hořava-Lifshitz gravity comparison}
\label{app:kh}

In this Appendix we illustrate the differences in the approaches used to implement Ho\v rava-Lifshitz gravity in CLASS-LVDM and \texttt{EFTCAMB}.

(1) The first key difference between CLASS-LVDM and \texttt{EFTCAMB} is the treatment of the background.
It is well known that the only effect of Ho\v rava-Lifshitz (khrononmetric) gravity on the homogeneous and isotropic
universe is the rescaling of the gravitational constant in the Friedman equation.
CLASS-LVDM uses the rescaled background densities defined via the Friedman equation
\begin{equation} 
\label{eq:classback}
H^2 = \sum_i \tilde{\rho}_i =  H_0^2\sum_{i} \tilde\Omega_i(z) \,,
\end{equation}
in which way the densities $\tilde{\rho}_i$ (correspondingly, 
$\tilde\Omega_i(z=0)$ - subject to input in CLASS-LVDM) are rescaled by $G_{cos}/G_N$,
and the flatness condition $\sum_{all\,species}\tilde\Omega_i(z=0)=1$ is satisfied automatically.

On the other hand, \texttt{EFTCAMB} uses the background densities defined via the gravitational constant in the Newtonian limit,
which is generically different from that appearing in the Friedman equation.
To be more precise, \texttt{EFTCAMB} solves the following Friedman equation 
\begin{equation} 
\label{eq:cambback}
H^2 = H_0^2\frac{G_{cos}}{G_{N} }\left(\sum_{dm,\,b,\,\gamma,\,\nu} \Omega_i(z) + \left[\Omega_{DE}^0 + \frac{G_N}{G_{cos}} -1\right] \right)\,,
\end{equation}
where $\Omega_{DE}^0$ is the present time DE density parameter. The fractional densities $\Omega_i(z=0)$ (subject to input in \texttt{EFTCAMB}) are therefore the ``bare" parameters.
The modification in the effective $\Omega_{DE}^0$ in the square brackets of \eqref{eq:cambback} is dictated by the requirement 
that the flatness condition ($\sum_{all\,species}\Omega_i =1 $) be satisfied at redshift zero \cite{Frusciante:2015maa}.\footnote{Note that one could redefine $\Omega_{DE}(z)$ to absorb all the modifications due to the rescaling of the gravitational constant into it. This would lead to a $\Omega_{DE}(z)$ dependent on the HL parameters plus a standard gravitational constant. This different convention would lead to the same cosmology as with our definitions (if all ``bare” fractional densities are suitably chosen), since the two descriptions are equivalent.}

To sum up, the background evolution in both codes is intrinsically different in the case $G_N\neq G_{cos}$,
which is why for the purposes of this paper we focused only on the parameters for which $G_N = G_{cos}$.

(2) The second difference is the definition of the matter power spectrum.
As explained in Refs.~\cite{Audren:2013dwa,Audren:2014hza},
in order to match the observations, 
the power spectrum in CLASS-LVDM is rescaled by the factor $(G_{cosm}/G_N)^2$.
This is to be contrasted with \texttt{EFTCAMB}, which uses the standard definition. 
Within our convention to study only the case $G_N = G_{cos}$, this difference becomes irrelevant.

(3) The third difference is in the normalization of the primordial power spectrum.
In order to isolate the LV effects from the standard cosmological parameters, 
in the CLASS-LVDM code by default the initial power spectrum 
of metric perturbations is normalized in a way to match the $\Lambda$CDM one for the same choice of $A_s$ regardless 
of values of the LV parameters.
This is not the same in \texttt{EFTCAMB}, where additionally to the background densities the initial power spectrum
also bears the dependence on the the extra parameters of Ho\v{r}ava/khrononmetric gravity.
Qualitatively, there is no difference between these two approaches.
For the purposes of this paper for each set of parameters we normalized the initial power spectra
to the same value in both codes.

(4) The fourth difference is in the initial conditions. CLASS-LVDM assumes the initial conditions for the khronon field 
corresponding to the adiabatic mode \cite{Audren:2013dwa,Audren:2014hza}. On the other hand, \texttt{EFTCAMB}  assumes for the initial conditions that DE perturbations are sourced by matter perturbations at a sufficiently early time so that the theory is close to General Relativity \cite{Hu:2014oga}. In order to take into account the difference in the initial conditions, only for this comparison in both codes we set the initial conditions as
\begin{equation}
\label{eq:piIC}
\pi(\tau_0)=0 \quad \text{and} \quad  \dot\pi(\tau_0)=0\,,
\end{equation}
where $\pi$ is the extra scalar degree of freedom (i.e.\ khronon). It is important to note that this choice correspond to an isocurvature mode that totally compensates the adiabatic one at the initial time.

\bibliography{EB_comparison_PRD}

\end{document}